\documentclass[onecolumn,12pt]{ieeetran}
\usepackage{bm}
\usepackage{amsmath}
\usepackage{subfigure}
\usepackage{algorithm}
\usepackage{algorithmic}
\usepackage{booktabs}
\usepackage{color}
\usepackage{verbatim}
\usepackage{setspace}
\usepackage[breaklinks,colorlinks,linkcolor=black,citecolor=black,urlcolor=black]{hyperref}
\usepackage{multirow}
\usepackage{ulem}
\usepackage{geometry}
\newcommand*{\blue}{\textcolor{black}}

\usepackage[]{graphicx}    
\newcommand{\ignore}[1]{}  

\ifCLASSINFOpdf
\else
\fi
\begin{document}

\title{Large region targets observation scheduling by multiple satellites using resampling particle swarm optimization}

\author{Yi Gu, Chao Han, Yuhan Chen, Shenggang Liu, and Xinwei Wang 
	\thanks{Yi Gu, Chao Han, and Shenggang Liu are with the School of Astronautics, Beihang University, Beijing, 100191, China (e-mail: guyi\_buaa@buaa.edu.cn; hanchao@buaa.edu.cn;  liushg@buaa.edu.cn).} 
	\thanks{Yuhan Chen is with the China Satellite Network Innovation Co., Ltd, Beijing, 100029, China (e-mail: chenyuhan@buaa.edu.cn).}
	\thanks{Xinwei Wang is with the Department of Transport and Planning, Delft University of Technology, Delft 2628 CD, The Netherlands (e-mail: xinwei.wang.china@gmail.com). (\textit{Corresponding authors}: Shenggang Liu, Xinwei Wang.)}}

\markboth{IEEE Transactions on Aerospace and Electronic Systems}%
{Shell \MakeLowercase{\textit{et al.}}: Bare Demo of IEEEtran.cls for IEEE Journals}

\maketitle

\begin{abstract}
	\noindent The last decades have witnessed a rapid increase of Earth observation satellites (\blue{EOSs}), leading to the increasing complexity of \blue{EOSs scheduling}.
	On account of the widespread applications of large region observation, this paper aims to address the EOSs observation scheduling problem for large region targets.
	A rapid coverage calculation method employing \blue{a} projection reference plane and \blue{a} polygon clipping technique is first developed. We then formulate a nonlinear integer programming model for the scheduling problem, where the objective function is calculated based on the developed coverage calculation method.
	A greedy initialization-based resampling particle swarm optimization (GI-RPSO) algorithm is proposed to solve the model. 
	The adopted greedy initialization strategy and particle resampling method contribute to generating efficient and effective solutions during the evolution process.
	In the end, extensive experiments are conducted to illustrate the effectiveness and reliability of the proposed method.
	Compared to the traditional particle swarm optimization and the widely used greedy algorithm, the proposed GI-RPSO can improve the scheduling result by 5.42\% and 15.86\%, respectively. 
	
\end{abstract}

\begin{IEEEkeywords}
	large region targets, multiple satellites, observation scheduling, resampling particle swarm optimization
\end{IEEEkeywords}
\IEEEpeerreviewmaketitle

\section{Introduction}

Earth observation satellites (EOSs) play a crucial role in acquiring images of specific areas of the Earth~\cite{wang2020agile}.
Taking advantages of space-based location and rapid revisit ability, EOSs can collect  information \blue{on} the Earth's surface conveniently by the observation payload. 
The last decades have seen a growing trend towards observation requests for the large region target, whose boundary size is much larger than the imaging width of EOSs. 
Thus  cooperative observation by multiple EOSs is necessary for  large region target observation, which is essential to environmental monitoring, crop survey, sea reconnaissance, and other practical areas~\cite{wang2021robust}.
The large region targets observation scheduling problem has attracted increasing attention due to its significant economic and social value.
Meanwhile, the growing number of EOSs~\cite{gu2021mission,wu2022ensemble} makes it possible for the cooperation observation by multiple EOSs.

\begin{figure}[htbp]
	\begin{center}
		\includegraphics[width=0.55\textwidth]{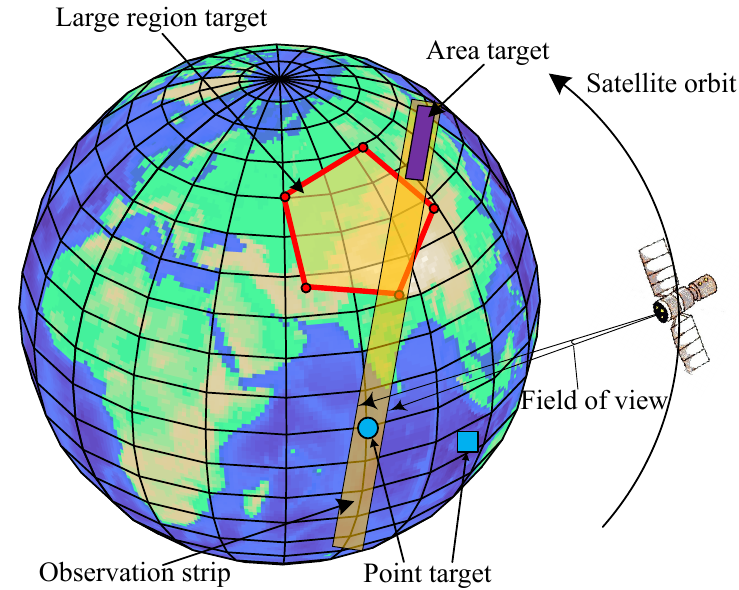}
		\caption{Schematic of \blue{the} imaging for different targets on the Earth.}
		\label{fig:Earth}
	\end{center}
\end{figure}

According to the relative size between the target and the field of view of the satellite, the observation targets on the Earth can be divided into point targets and area targets.
To further distinguish from the area target which can be fully observed during only one pass of satellite~\cite{chang2021integrated}, we define the large region target which can only be completely observed by  multiple EOSs cooperative observations.
Figure~\ref{fig:Earth} \blue{shows the} satellite observation for point and area targets on the Earth.
An EOS could generate a long observation strip whose width depends on the altitude and the field of view of the satellite.
Generally, the size of point targets and the small area target is less than the width of the observation strip and they can be successfully observed during one imaging.
On the contrary, the large region target enclosed by the red boundary can not be covered by the observation strip depicted in Figure~\ref{fig:Earth}.
This study will focus on the large region targets observation scheduling problem by multiple EOSs.

\subsection{Related work}

\begin{table*}[htbp]
	\centering
	\footnotesize
	\caption{Overview of the EOS \blue{scheduling} with respect to region target.}
	\begin{tabular}{lp{4.19em}ccp{4.19em}cc}
		\toprule
		\multicolumn{1}{c}{\multirow{2}[4]{*}{Satellite}} & \multicolumn{6}{c}{The classification of observation targets} \\
		\cmidrule{2-7}          & \multicolumn{3}{c}{Single region} & \multicolumn{3}{c}{Multiple regions} \\
		\midrule
		Single EOS & \multicolumn{3}{p{12.57em}}{Exact~\cite{lemaitre1997daily,mancel2003complex}, Heuristic~\cite{walton1993models,du2018area}} & \multicolumn{3}{c}{$-$} \\
		Multiple EOSs & \multicolumn{3}{p{12.57em}}{MOEA~\cite{wang2007approach,niu2018satellite}, Heuristic~\cite{zhu2010satellite,chen2012multi,perea2015swath,he2020balancing}} & \multicolumn{3}{p{12.57em}}{MOEA~\cite{li2018multiobjective,chen2020multi}, Heuristic~\cite{WangReinelt-50,zhu2019three,kim2015mission,kim2020optimal,xu2020multi,zhibo2021multi}} \\
		\bottomrule
	\end{tabular}%
	\label{tab: intro}%
\end{table*}%

A considerable amount of literature has been published on the point \blue{targets} observation scheduling problem over the past decades.
\blue{Existing algorithms to solve the Earth observation scheduling problem can be divided into three types:} the exact algorithm~\cite{LemaitreVerfaillie-269,WangDemeulemeester-4,wang2019robust}, heuristic~\cite{wang2016scheduling,liu2017adaptive,peng2019agile,wang2019scheduling}, and multi-objective evolutionary algorithm (MOEA)~\cite{du2019moea}.
Although region targets can be decomposed into point targets, the scale of the scheduling problem would increase dramatically with respect to large region targets.
Meanwhile, the dramatic increase in computational complexity  prevents the application of the observation scheduling method for point targets in the scheduling problem of observing large region targets.

Existing studies on the region target observation scheduling problem by EOSs are comparatively limited.
In order to make a clear description of the relevant literature, we summarize previous studies in Table~\ref{tab: intro} in line with applied scheduling algorithms. 
The single satellite scheduling problem for a region target was firstly addressed in~\cite{walton1993models}, where a set covering model was established.
Lema\^itre and Verfaillie~\cite{lemaitre1997daily} introduced \blue{an} integer linear programming for a single agile satellite scheduling problem and solved the constraint programming model with \blue{a} commercial solver.
Mancel \blue{and Lopez}~\cite{mancel2003complex} adopted \blue{a} column generation algorithm to solve the integer programming model for scheduling the PLEIADES satellite of France.
The observation coverage time was optimized by the particle swarm optimization (PSO) algorithm hybridized with a differential evolutionary algorithm.
To accomplish the observation for an area target by a single satellite, Du $et\ al$.~\cite{du2018area} transformed the original problem \blue{into} a path planning problem for visiting nodes and developed an improved ant colony algorithm to solve it.
\blue{Vazquez $et\ al$.~\cite{vazquez2014resolution} proposed an automated satellite antenna assignment algorithm for  observation data transmission.}

\blue{Due to the great advantage of multiple EOSs cooperation, most research has been conducted on the multiple EOSs scheduling problem.} 
\blue{For observing a single region target}, Wang $et\ al$.~\cite{wang2007approach} constructed a multi-objective optimization model and adopted a strength Pareto evolutionary algorithm to obtain the scheduling result.
For disaster emergency response, Niu $et\ al$.~\cite{niu2018satellite} solved the satellite scheduling problem of a large region task by using a multi-objective genetic algorithm.
Four optimization objectives including the imaging finish time, the coverage, the mean spatial resolution, and the average slewing angle have been taken into consideration.
The 2008 Wenchuan earthquake was taken as a real disaster scenario to evaluate the performance of the algorithm.
For the rapid observation of a target area \blue{in} an emergency, Zhu $et\ al$.~\cite{zhu2010satellite} \blue{have} taken the satellite scheduling and orbital transfer \blue{problems} into consideration, simultaneously.
In order to maximize the observation coverage, Chen $et\ al$.~\cite{chen2012multi} proposed a hybrid optimization method based on the PSO and genetic algorithm.
Perea $et\ al$.~\cite{perea2015swath} modeled the scheduling problem as a set covering problem and presented an adaptive search algorithm to solve it.
He $et\ al$.~\cite{he2020balancing} constructed an integer programming model for the area target scheduling problem, which aims at maximizing the coverage ratio while minimizing the response time by considering the weighted sum.
A balanced heuristic has been presented to solve the model and the constant-factor performance guarantee can be provided.

As for the multi-region observation, Li $et\ al$.~\cite{li2018multiobjective} introduced the preference incorporation to the region targets observation scheduling problem and established a multi-objective optimization model considering the profit, quality, and timeliness simultaneously.
In order to maximize the target area coverage and minimize the observation resource utilization, Chen $et\ al$.~\cite{chen2020multi} adopted a non-dominated sorting genetic algorithm to solve the model. 
Wang $et\ al$.~\cite{WangReinelt-50} established a nonlinear model and presented a heuristic for solving the disaster monitoring observation scheduling problem by four satellites.
For the polygon region observation request, Zhu $et\ al$.~\cite{zhu2019three} proposed a three-phase solution method that integrates with a dynamic greedy algorithm and a tabu search algorithm.
Moreover, the scheduling problem for the synthetic aperture radar satellite, which owns a specific field of view, has been researched for area observing~\cite{kim2015mission,kim2020optimal}.
Focusing on large region target observation, Xu $et\ al$.~\cite{xu2020multi} and E $et\ al$.~\cite{zhibo2021multi} developed improved genetic algorithms to solve the multi-satellite scheduling problem, aiming at improving the summed observation profit. 

The calculation \blue{of observation coverage} is a significant component in the observation scheduling of large region targets.
Two calculation methods are mainly utilized to determine the coverage in previous studies.
The grid discretization-based area calculation (GDAC) method is widely adopted owing to its good performance in precision~\cite{xu2020multi,zhibo2021multi}.
\blue{While concerning large region targets, we have to strike a balance between the computational efficiency and the calculation accuracy of the GDAC.}
Meanwhile, the polygon clipping technique for plane polygon intersection operation has also been utilized to conduct the coverage analysis~\cite{wang2022versatile}. 
It \blue{is worth pointing} out that a general projection from a spherical to a plane, \blue{serving} as the basis of clipping, will result in relatively large deformation.

\subsection{Objective and contributions}

Although the above research has been carried out on area target observation by EOSs, few studies pay sufficient attention to the large region target except for~\cite{niu2018satellite,chen2020multi,xu2020multi,zhibo2021multi}.
A possible explanation might be that there exist extensive candidate tasks derived \blue{from} a large region target, which makes the coverage calculation and scheduling process more difficult.
However, there still exist two challenges for large region targets observation scheduling.
First, existing studies generally calculate the plane intersection between the large region and observation strips instead of \blue{the} spherical intersection.
The equidistant cylindrical projection in~\cite{zhibo2021multi} is reasonable when the region target is small.
With respect to the large region target or high latitude region, the area boundary would be greatly deformed during the projection between plane and sphere.
This results in inaccurate calculation of the intersection, and even affects \blue{the} successful execution of the observation.
Second, the approximate coverage calculation by counting the number of grid points is relatively time-consuming.
The calculation complexity would be sharply increased when larger region targets are required to observe.

The large region targets observation scheduling is additionally complicated by extensive observation strips and the considerable solution space.
This study, therefore, dedicates to investigating a rapid and accurate coverage calculation method and then establishing an efficient scheduling algorithm for the large region target observation by multiple EOSs.
Unlike existing studies, a coverage calculation method based on \blue{a} polygon clipping technique~\cite{max2005computer} is proposed to substitute the GDAC method. 
To eliminate the calculation bias between a plane and a spherical surface, a projection reference plane is defined for each large region target and the center projection method is adopted.
The precise intersection point between the region target and satellite strips can then be obtained through the proposed projection transformation and inverse transformation.
\blue{The presented projection reference plane integrated polygon clipping technique can be denoted as the amended polygon clipping technique (APCT) in this study.} 
On the basis of the proposed coverage calculation method, a nonlinear integer programming model is constructed for maximizing the total observation area and satisfying engineering constraints. 
Given  the  multiple EOSs scheduling has been proved  NP-hard~\cite{zhibo2021multi,xhafa2021optimisation}, we propose an improved resampling particle swarm optimization (RPSO) algorithm to solve the established model.
The algorithm based on particle searching  has been demonstrated as well-performance in the large-scale optimization problem~\cite{wang2018coverage}.
The greedy initialization strategy is designed to enhance the convergence speed while the presented resampling method has \blue{the} potential to obtain better solutions.
Moreover, a particle reconstruction method has been adopted to handle several engineering constraints for practical applications.

The main contributions of this study can be summarized as twofold: (1) We propose an accurate coverage calculation method based on the specially designed projection reference plane and polygon clipping technique, which ensures efficient objective function calculation of the scheduling problem. (2) An improved RPSO scheduling algorithm, integrated with the greedy initialization strategy and the particle reconstruction method, has been developed to enhance the large region targets scheduling result.

The remaining part of this paper proceeds as follows.
Section~\ref{sec: coverage} describes the region coverage calculation method. 
In Section~\ref{sec: model}, 
the nonlinear integer mathematical model is constructed in line with necessary assumptions and notations.
Section~\ref{sec: IRPSO} presents the overall  framework of RPSO, followed by \blue{a} detailed design of each algorithm component.
Experiment setting and simulation results are provided in Section~\ref{sec: result}.
Finally, the paper ends with  conclusions in Section~\ref{Sec: conclu}. 
\section{Coverage calculation method based on polygon clipping}
\label{sec: coverage}

With respect to the large region targets observation problem, we aim to maximize the total coverage for the regions to be observed~\cite{zhibo2021multi}.
Efficient and accurate coverage calculation of each area is thus essential. 
In this section, we first describe the observation scheduling problem for large region targets, and then define the projection reference plane. 
Based on the introduced polygon clipping technique, the coverage calculation method can be finally provided.
\subsection{Problem description}

\begin{figure}[htbp]
	\begin{center}
		\includegraphics[width=0.55\textwidth]{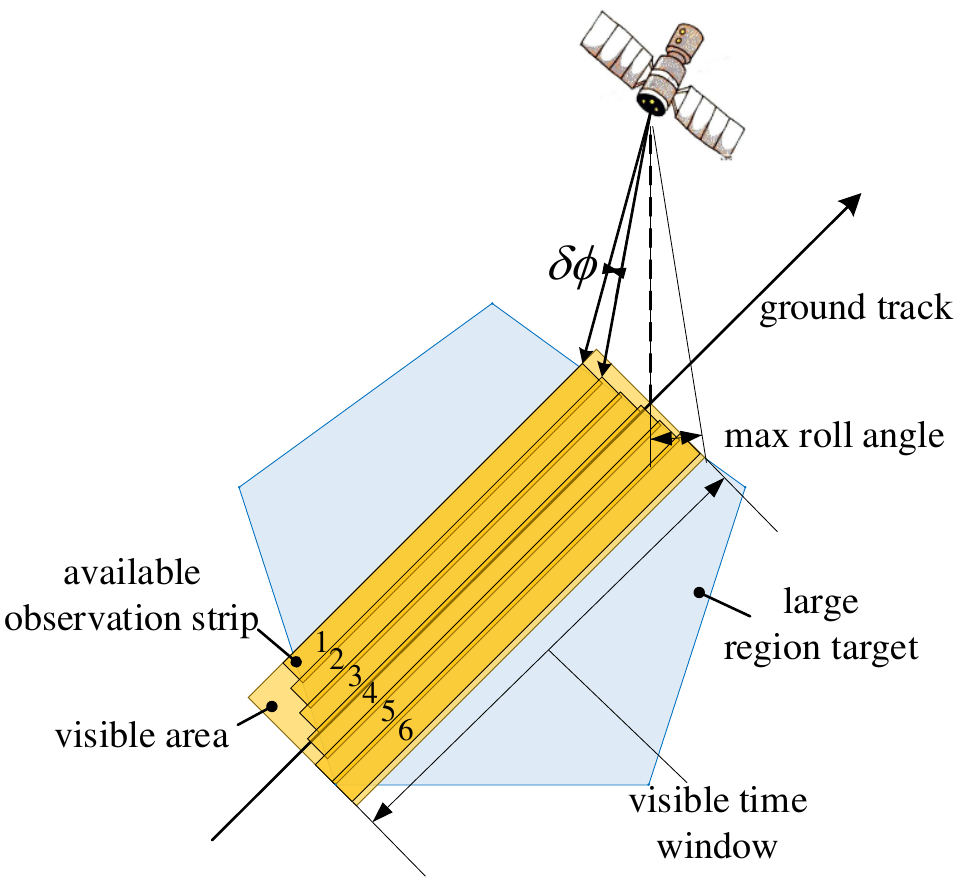}
		\caption{Schematic of satellite imaging and observation strips.}
		\label{fig:Satellite}
	\end{center}
\end{figure}

As  seen in Figure~\ref{fig:Satellite}, an EOS is flying over a large region target, and the visible time window for the target can be determined according to the area boundaries and the satellite maneuverability.
Generally, the field of view angle of the payload is limited, but the satellite can adjust the orientation of the payload through attitude maneuvering so as to increase the observable area. 
Taking the satellite rolling ability into consideration, the visible area along with the satellite ground track can be uniquely defined. 
The boundary of the visible area along the ground track corresponds to the visible start time and end time, respectively.
Meanwhile, \blue{the} boundaries of the visible area perpendicular to the ground track are tangent to the boundary of the region target.
Available observation strips are discretely distributed within the visible area and the roll angle interval between two adjacent observation strips is denoted as $\delta \phi$.
Note that only one observation strip can be collected during one passing of a satellite.
As shown in Figure~\ref{fig:Satellite}, each observation strip owns a specific observation duration which corresponds to varying degrees of coverage.
Thus the selection \blue{of} an observation strip plays a key role in the scheduling problem.

Note that although Figure~\ref{fig:Satellite} is displayed on a plane surface for the sake of simplification, the division of observation strips and coverage calculation actually are performed on the spherical surface.
By dividing the observation strip according to a certain time duration, each independent observation strip can be approximated as a spherical polygon on the surface of the Earth.
Meanwhile, large region \blue{targets} in this study \blue{are defined} as spherical \blue{polygons} of the Earth.
For \blue{regions} with other different definitions, \blue{they can generally} be  approximated as spherical polygons through \blue{discretization}.
Then the coverage calculation can be transformed into the problem of intersection and merging of spherical polygons.
The intersection strip represents the intersection area of the observation strip and the large region target.
Based on the polygon clipping technique, boundary points after merging different intersection strips can be obtained to support the coverage calculation. 

\subsection{Projection reference plane} 
Calculating the coverage of satellite observation strips plays a critical role in the \blue{observation} scheduling of large region targets.
There exist two \blue{main} methods for computing intersection strips: plane \blue{polygons} and spherical \blue{polygons}.
Owing to the fact that the intersection computing based on spherical \blue{polygons} is time-consuming and inefficient~\cite{xu2020multi}, the intersection on plane \blue{polygons} is utilized in this study.

To reduce the distortion during the projection transformation from sphere to plane, a projection reference plane is defined for each large region target.
This means that observation strips and region targets will be projected on the reference plane for intersection and merging computing.
Meanwhile, a central projection rule which takes the Earth center as the projection reference point is adopted. Thus the spherical polygon boundary can be accurately projected as a straight line on the projection plane.

\begin{figure}[htb]
	\begin{center}
		\includegraphics[width=0.55\textwidth]{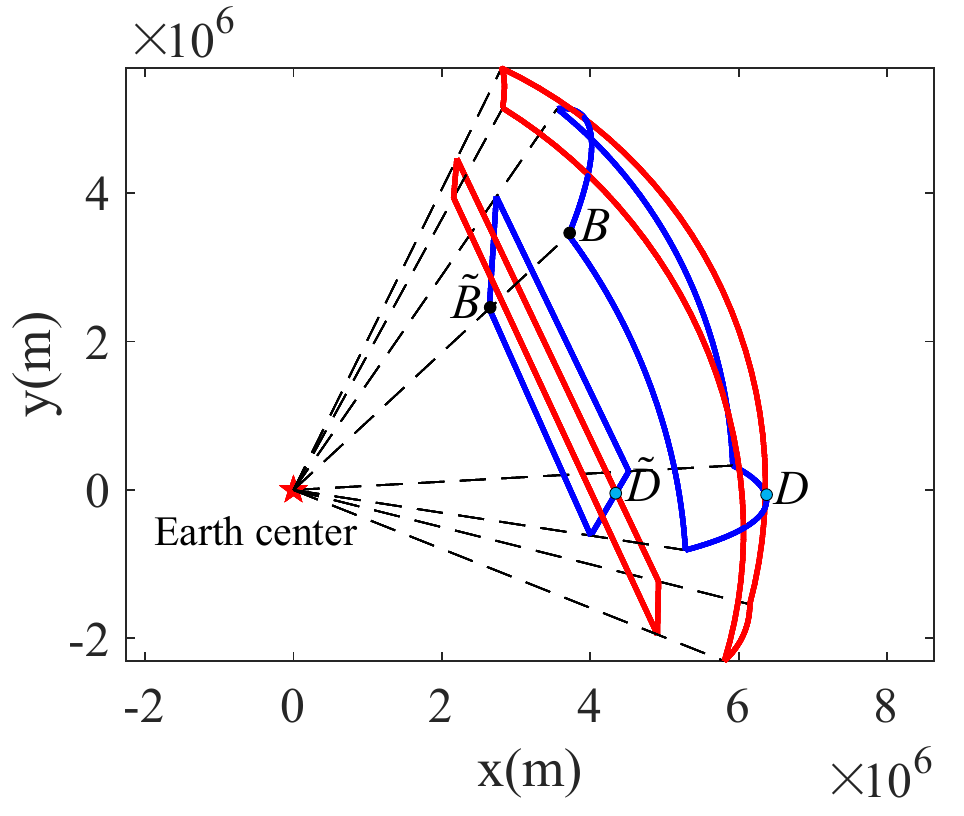}
		\caption{Schematic of projection from Earth surface to reference plane.}
		\label{fig:Polygon}
	\end{center}
\end{figure}
As shown in Figure~\ref{fig:Polygon}, two spherical polygons in different colors are projected onto the reference plane. The blue polygon indicates a large region target while the red one could denote an observation strip.
Points $B$ and $D$ represent an area boundary point and an intersection of two polygons, respectively.
Note that all intersection points could be determined accurately in this way.
The projection reference plane should be determined as close to the large region target as possible to reduce deformation.
A general framework to select the projection reference plane for each large region target is proposed as follows.
\begin{enumerate}
	\item \blue{The} set of boundary points of a large region target is denoted as $P$, and $|P|$ indicates the set size. 
	Coordinates of each boundary point in the Earth-Centered Earth-Fixed (ECEF) coordinate system can be denoted as $(x_{p}, y_{p}, z_{p}), p\in P$.
	The arithmetic average center of these boundary points can be referred to as $C$.
	Denote the vector from Earth center to point $C$ as $\bm{R}_{c}=(x_{c}, y_{c}, z_{c})$, which can be formulated as 
	\begin{align}
	\left\{ {\begin{array}{*{20}{c}}
		x_{c} = \sum\limits_{p\in{P}} x_{p}/|P|\\ 
		y_{c} = \sum\limits_{p\in{P}} y_{p}/|P|\\
		z_{c} = \sum\limits_{p\in{P}} z_{p}/|P|
		\end{array}} \right. \label{eq: xc}
	\end{align}
	\item 
	To describe the reference plane, a cartesian coordinate system $S_{c}$ is established with point $C$ as the origin.
	Three unit basis vectors of $S_{c}$ can be denoted as 
\begin{align}
	\left\{ {\begin{array}{*{20}{c}}
			\bm{i}_{c} = \dfrac{\bm{z}_{f} \times \bm{R}_{c}}{\|\bm{z}_{f} \times \bm{R}_{c}\|}\\
			\bm{j}_{c} = \bm{k}_{c} \times \bm{i}_{c}\\
			\bm{k}_{c} = \bm{R}_{c}/\|\bm{R}_{c}\| 
	\end{array}} \right. \label{eq: vector}
\end{align}
	where $\bm{z}_{f}$ indicates the unit vector of the $z$-axis in ECEF. 
	\item The projection reference plane can be set as $z = \|\bm{R}_{c}\|$ in $S_{c}$.
	For arbitrary spherical polygon boundary point $B$, denote  $\bm{R}_{B}=(x_{B}, y_{B}, z_{B})$ as the vector from Earth center to point $B$ in ECEF coordinate system.
	Then, point $B$ can be projected onto the reference plane as $(x_{B}, y_{B}, z_{B}) \rightarrow (\widetilde{x}_{B},\widetilde{y}_{B})$, and the projection coordinate in $S_{c}$ can be calculated as
	\begin{align}
	\left\{ {\begin{array}{*{20}{c}}
		 \widetilde{x}_{B} = \dfrac{\bm{i}_{c} \cdot \bm{R}_{B}}{\|\bm{i}_{c} \cdot \bm{R}_{B}\|} \cdot \|\bm{R}_{c} \|\\
		\widetilde{y}_{B} = \dfrac{\bm{j}_{c} \cdot \bm{R}_{B}}{\|\bm{j}_{c} \cdot \bm{R}_{B}\|} \cdot \|\bm{R}_{c} \|
		\end{array}} \right. \label{eq: pro}
	\end{align}
	Then the intersection computing based on the projection reference plane can be conducted utilizing the polygon clipping technique.
\end{enumerate}

\subsection{Polygon clipping technique}

Recent decades have seen an extensive application of polygon clipping in industrial manufacturing and computer-aided design~\cite{martinez2009new,simonson2010industrial}.
In this study, both the observation strip of EOSs and the large region target are not restricted as the convex polygon.
Therefore, the general efficient polygon clipping technique~\cite{vatti1992generic,max2005computer} which supports convex and concave polygons simultaneously is utilized to conduct the intersection and merging computing.
The C++ version of \blue{the} polygon clipping code can be obtained from the website: http://www.angusj.com.

\begin{figure}[htbp]
	\begin{center}
		\includegraphics[width=0.55\textwidth]{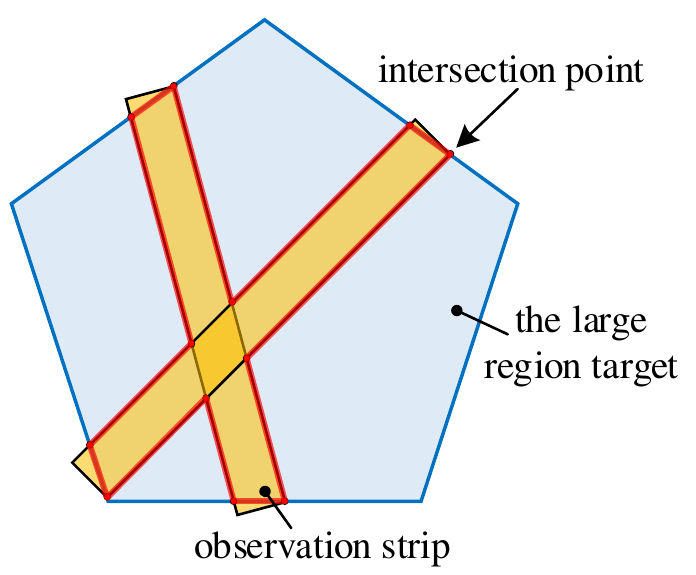}
		\caption{Schematic of \blue{the} intersection and merging of observation strips.}
		\label{fig:Intersection}
	\end{center}
\end{figure}

The utilization of the polygon clipping technique mainly consists of two stages. 
As depicted in Figure~\ref{fig:Intersection}, all intersection points between observation strips and large region targets on the projection reference plane can be obtained in the first stage. 
Intersection points among all observation strips could be calculated in the second stage.
Then the union of observation strips can be generated, which is the area with \blue{the} red color boundary in Figure~\ref{fig:Intersection}.
Meanwhile, coordinates of all intersection points in the reference coordinate system $S_{c}$ can be determined.  

\subsection{Strip coverage calculation}
In order to transform the result of intersection and merging operation on the plane to spherical, it is required to perform an inverse transformation. 
After generating boundary points of the union observation strip on the reference plane, an inverse transformation should be conducted for arbitrary intersection point $\widetilde{D}$ in Figure~\ref{fig:Polygon}.
It is not difficult to know that point $D$ is the result of \blue{the} inverse transformation, which is the intersection between the \blue{Earth's} surface and the line crossing both the \blue{Earth's} center and point $\widetilde{D}$.
Denote coordinates of point $\widetilde{D}$ on the reference plane as $(\widetilde{x}_{D}, \widetilde{y}_{D})$, and the vector from the Earth center to point $\widetilde{D}$ can be described as  
\begin{equation}
	\bm{R}_{\widetilde{D}} = \widetilde{x}_{D} \bm{i}_{c} + \widetilde{y}_{D} \bm{j}_{c} + \|\bm{R}_{c} \| \bm{k}_{c}
\end{equation}

Projected coordinates of $\bm{R}_{\widetilde{D}}$ in ECEF can be denoted as $[x_{Df}, y_{Df}, z_{Df}]^{T}$, and corresponding coordinates of point $D$ on the Earth can be formulated as
\begin{align}
	\left\{ {\begin{array}{*{20}{c}}
			x_{D} = x_{Df}/Nor \\
			y_{D} = y_{Df}/Nor\\
			z_{D} = z_{Df}/Nor
	\end{array}} \right. 
\end{align}
\begin{equation}
	Nor = \sqrt {{{\left( {\frac{{{x_{Df}}}}{{{R_{E}}}}} \right)}^2} + {{\left( {\frac{{{y_{Df}}}}{{{R_{E}}}}} \right)}^2} + {{\left( {\frac{{{z_{Df}}}}{{{R_{E}}}}} \right)}^2}}
\end{equation}
where $R_{E}$ indicates the equatorial radius of the Earth.
Subsequently, the calculation method of spherical polygon area~\cite{chamberlain2007some} can be employed to determine the strip coverage.



\section{Large Region Targets Observation Scheduling by EOSs}
\label{sec: model}

\subsection{Assumptions}
This study focuses on solving the observation scheduling problem of large region targets by multiple EOSs.
The following necessary assumptions and simplifications are adopted for practical applications.
\begin{enumerate}
	\item This study takes large region targets into account which can not be completely observed in one pass of a satellite. 
	\item The EOS mentioned in this paper refers to the non-agile satellite without pitching maneuverability. 
	\item \label{assump: 3} Only one observation can be conducted by each satellite at the same time and the observation task cannot be interrupted. 
	\item \label{assump: 4} The satellite attitude remains unchanged during the observation to ensure imaging quality.
	\item Supposing that there exist enough ground stations in the scenario, the data download, and instruction upload fall outside of the scope of our study. 
\end{enumerate}

\subsection{Notations}
\begin{table}[htbp]
	\centering
	\caption{Notations.}
	\footnotesize
	\begin{tabular}{lp{35.25em}}
		\toprule
		\multicolumn{2}{l}{\textbf{Targets}} \\
		$T$   & Set of large region targets, $T = \{ 1,...,|T|\}$ and $|T|$ is the target set size \\
		$i$   & Target index, $i \in T$ \\ 
		$i_{s}$ & Target index of denoting the successor observation of $i$, $i_{s} \in T\cup \{|T|+1\}$, in which $|T|+1$ is a dummy target \\
		$i_{1}, i_{2}$ & Target index, $i_{1} \in T, i_{2} \in T$ \\
		$A_{i}$ & The area of large region target $i$, $i\in T$ \\
		\textbf{Satellies \& Orbits} &  \\
		$S$   & Set of satellites, $S = \{ 1,...,|S|\}$ and $|S|$ is the satellite set size \\
		$j$   & Satellite index, $j \in S$ \\
		$O_{j}$ & Set of orbits of satellite $j$, $O_{j} = \{ 1,...,|O_{j}|\}$ and $|O_{j}|$ is the orbit set size, $j\in S$ \\
		$k, k_{1}, k_{2}$ & Orbit index, $k, k_{1}, k_{2}\in O_{j}$ \\
		$M_{jk}, E_{jk}$ & Memory and energy capacity of satellite $j$ on orbit $k$, $j\in S, k\in O_{j}$ \\
		$et_{jk}$ & Energy consumption for each unit angle of attitude transition of satellite $j$ on orbit $k$, $j\in S, k\in O_{j}$ \\
		$mo_{jk},eo_{jk}$ & Memory and energy consumption for unit time observation of satellite $j$ on orbit $k$, $j\in S, k\in O_{j}$ \\
		$[VS_{ijk},VE_{ijk}]$ & Visible time window of large region target $i$ by satellite $j$ on orbit $k$, $i\in T, j \in S, k\in O_{j}$ \\
		$\delta \phi$ & Real parameter, indicting the discrete granularity interval of roll angle \\
		\textbf{Observation strips} &  \\
		$L_{ijk}$ & Set of feasible observation strips, $L_{ijk} = \{ 1,...,|L_{ijk}|\}$ and $|L_{ijk}|$ is the number of observation strips of target $i$ by satellite $j$ on orbit $k$, $i\in T, j\in S, k\in O_{j}$ \\
		$l, l_{1}, l_{2}$ & Strip index,  $l\in L_{ijk}$, $l_{1}\in L_{i_{1}jk_{1}}$, $l_{2}\in L_{i_{2}jk_{2}}$, $i,i_{1},i_{2}\in T$, $j\in S$, $k,k_{1},k_{2}\in O_{j}$ \\
		$l_{s}$ & Strip index of denoting the successor observation of $l$,  $l_{s}\in L_{i_{s}jk}$, $i_{s} \in T\cup \{|T|+1\}, j\in S, k\in O_{j}$ \\
		$[OS_{ijk}^{l},OE_{ijk}^{l}]$ & Observation time window of strip $l$ within target $i$ by satellite $j$ on orbit $k$, $l\in L_{ijk}, i\in T, j\in S, k\in O_{j}$ \\
		$od_{ijk}^{l}$ & The observation duration of strip $l$ within target $i$ by satellite $j$ on orbit $k$, $l\in L_{ijk}, i\in T, j\in S, k\in O_{j}$ \\
		$st_{j}^{l_{1}l_{2}}$ & Attitude transition time between observation strips $l_{1}$ and $l_{2}$ of satellite $j$, $j\in S$, $l_{1}\in L_{i_{1}jk_{1}}$, $l_{2}\in L_{i_{2}jk_{2}}$ \\
		$se_{ijk}^{ll_{s}}$ & Energy consumption in the transition from strip $l$ to $l_{s}$ of satellite $j$ on orbit $k$, $i\in T, j\in S, k\in O_{j}$, $l\in L_{ijk}, l_{s}\in L_{i_{s}jk}$ \\
		\multicolumn{2}{l}{\textbf{Decision variable}} \\
		$x_{ijk}^{l}$ & Binary decision variable. $x_{ijk}^{l} = 1$ if strip $l$ within target $i$ is scheduled to be observed by satellite $j$ on orbit $k$, otherwise $x_{ijk}^{l} = 0$, $l\in L_{ijk}, i\in T, j\in S, k\in O_{j}$ \\
		\bottomrule
	\end{tabular}%
	\label{tab: Notations}%
\end{table}%

For ease of reference, some notations in this study are summarized in Table~\ref{tab: Notations}.
Let $T$ and $S$ be the set of large region targets and satellites, respectively.
For each satellite $j\in S$, $O_{j}$ indicates the set of orbits of satellite $j$ during the given scenario.
Each orbit $k\in O_{j}$ is associated with five parameters: $M_{jk}$ indicting the maximum memory capacity, $E_{jk}$ denoting the maximum energy capability, $et_{jk}$ denoting the energy consumption for a unit angle of attitude transition, $mo_{jk}$ and $eo_{jk}$ representing the memory and energy consumption for unit time observation, respectively.
Moreover, the visible time window $[VS_{ijk},VE_{ijk}]$ denotes visible start and end time of large region target $i$ by satellite $j$ on orbit $k$, corresponding to a specific visible area.
A visible area can be obtained according to the intersection of large region target $i$ and the satellite's field of view within $[VS_{ijk},VE_{ijk}]$.
Then the visible area can be divided into several observation strips relying on $\delta \phi$, namely the discrete interval of \blue{the} roll angle.
Let $L_{ijk}$ \blue{denote} the set of available observation strips formulated by the division of the visible area.
The observation time window $[OS_{ijk}^{l},OE_{ijk}^{l}]$ can be calculated for each strip $l\in L_{ijk}$.
The observation duration of strip $l$ by satellite $j$ on orbit $k$ is described as $od_{ijk}^{l} = OE_{ijk}^{l} - OS_{ijk}^{l}$.

When a satellite accomplishes an observation for strip $l_{1}\in L_{i_{1}jk_{1}}$, an attitude transition process is needed before an observation for strip $l_{2}\in L_{i_{2}jk_{2}}$.
Notice the required transition time $st_{j}(i_{1},k_{1},l_{1},i_{2},k_{2},l_{2})$ from $l_{1}$ to $l_{2}$ is also related to target index $i_{1}, i_{2}$ and orbit index $k_{1}, k_{2}$.
We denote the attitude transition time between \blue{adjacent} observation strips as $st_{j}^{l_{1}l_{2}}$ for simplicity afterward.
Taking the attitude stabilization process into account~\cite{han2020simulated,wang2018modified}, the attitude transition time can be formulated as 
\begin{equation}
	st_{j}^{l_{1}l_{2}} = \displaystyle{\frac{|\phi_{i_{2}}^{l_{2}}-\phi_{i_{1}}^{l_{1}}|}{v_{j}^{roll}}}{\rm{ + }}\left\{ {\begin{array}{*{20}{c}}
			5&{}&{|\phi_{i_{2}}^{l_{2}}-\phi_{i_{1}}^{l_{1}}| \le 15^{\circ}}\\
			{10}&{}&{15^{\circ} < |\phi_{i_{2}}^{l_{2}}-\phi_{i_{1}}^{l_{1}}| \le 40^{\circ}}\\
			{15}&{}&{40^{\circ} < |\phi_{i_{2}}^{l_{2}}-\phi_{i_{1}}^{l_{1}}| }
	\end{array}} \right.
\end{equation}
where $v_{j}^{roll}$ represents the attitude maneuvering angular velocity of the roll axis for satellite $j$, $\phi_{i_{2}}^{l_{2}}$ and $\phi_{i_{1}}^{l_{1}}$ indicate roll angles for observing strips $l_{1}$ and $l_{2}$, respectively.
The attitude maneuvering process from successive observation strips $l$ to $l_{s}$ by satellite $j$ on orbit $k$ consumes energy $se_{jk}(i,i_{s},l,l_{s})$, which can be simplified as $se_{ijk}^{ll_{s}}$.
In order to maintain satellite energy security, we assume that the roll angle of EOSs at the beginning of each orbital period is zero when calculating the total energy consumption on each orbit.
If target $i$ is the initial or ending observation on orbit $k$, therefore, corresponding energy consumption during attitude transition can be defined as ${et_{jk} \cdot |\phi_{i}^{l}|}$.
The energy consumption during each transition process can be calculated as
\begin{equation}
	se_{ijk}^{ll_{s}} = \left\{ {\begin{array}{*{20}{c}}
			{et_{jk} \cdot |\phi_{i}^{l}|}&{}&{i \text{ is the initial observation}}\\
			{et_{jk} \cdot |\phi_{i}^{l}|}&{}&{i\text{ is the ending observation}}\\
			{et_{jk} \cdot |\phi_{i_{s}}^{l_{s}}-\phi_{i}^{l}|}&{}&{\text{otherwise}}
	\end{array}} \right. \label{eq: maneuver}
\end{equation}
where $\phi_{i}^{l}$ and $\phi_{i_{s}}^{l_{s}}$ represent observation roll angles for strips $l\in  L_{ijk}$ and $l_{s}\in L_{i_{s}jk}$, respectively.

\subsection{Mathematical model} 
\label{subsec: model}

Based on \blue{the} above notations and equations, the mathematical model for the large region targets observation scheduling problem by multiple EOSs can be constructed as follows.
\begin{align}
\text{max}\quad& \sum\limits_{i\in{T}}  \frac{\xi_{i}}{A_{i}} \cdot f \Big ( \blue{\bigcup_{{j\in{S},k\in{O_{j}},l\in{L_{ijk}}}} x_{ijk}^{l}}\Big ) \label{ObjFuncNew}
\intertext{subject to} & \sum\limits_{l\in{L_{ijk}}} x_{ijk}^{l} \le 1 & \forall i\in T, j\in S, k\in O_{j}  \label{Cons1}	\\
& \sum\limits_{i\in{T}} \sum\limits_{l\in{L_{ijk}}} (x_{ijk}^{l} \cdot od_{ijk}^{l} \cdot mo_{jk}) \le M_{jk} & \forall j\in S, k\in O_{j} \label{Cons3} \\
& \sum\limits_{i\in{T}} \sum\limits_{l\in{L_{ijk}}} x_{ijk}^{l} \cdot (od_{ijk}^{l} \cdot eo_{jk} + se_{ijk}^{ll_{s}}) \le E_{jk} & \forall j\in S, k\in O_{j} \label{Cons4} \\ 
& x_{ijk}^{l} \in \{0,1\} & \forall i\in T, j\in S, k\in O_{j} \label{Cons5}
\end{align}
\begin{equation}
\begin{aligned}
\big\{x_{i_{1}jk_{1}}^{l_{1}} \cdot x_{i_{2}jk_{2}}^{l_{2}} \cdot(OS_{i_{2}jk_{2}}^{l_{2}} - &  OE_{i_{1}jk_{1}}^{l_{1}} - st_{j}^{l_{1}l_{2}})  \ge 0 | \text{if } OS_{i_{2}jk_{2}}^{l_{2}} \ge OE_{i_{1}jk_{1}}^{l_{1}} \big\} \\
& \forall i_{1},i_{2} \in T, j\in S, k_{1},k_{2}\in O_{j},l_{1}\in L_{i_{1}jk_{1}}, l_{2}\in L_{i_{2}jk_{2}} \label{Cons6}
\end{aligned}
\end{equation}

\begin{itemize}
	\item The objective function~\eqref{ObjFuncNew} is to maximize the total area of the observed region.
    \blue{The collection of scheduled observation strips within target $i$ is denoted as $\Big ( \blue{\bigcup_{{j\in{S},k\in{O_{j}},l\in{L_{ijk}}}} x_{ijk}^{l}}\Big )$, and the total area can be calculated by an implicit function $f$, which corresponds to the coverage calculation} method illustrated in Section~\ref{sec: coverage}.
	\blue{A} coefficient factor $\xi_{i}$ is defined to balance the area difference of different regions, which can be described as
	\begin{equation}
	\xi_{i} = \frac{A_{i}}{\sum\limits_{i\in{T}} A_{i}} \times 100 \label{eq: xi}
	\end{equation}
	Thus the maximum value of the objective function~\eqref{ObjFuncNew} can be normalized as 100 regardless of the number of \blue{regional} targets in the scenario.
	\blue{Note we can also prioritize large region targets by introducing a weight coefficient $w_{i}$ for each large region as $\xi_{i}^{'}=\xi_{i} \cdot w_{i}$. We may also vary the objective function, and set a hard constraint for observing an entire large region target, which is out of the scope of this work.}
	\item Constraints~\eqref{Cons1} regulate that for each region target, at most one available observation strip can be observed on every satellite orbit, which is guaranteed by assumptions~\ref{assump: 3} and~\ref{assump: 4}.
	\item Constraints~\eqref{Cons3} restrict that the memory consumption for observations is always within the memory capacity on each orbit.
	\item Constraints~\eqref{Cons4} indicate that the total energy consumption from target observing and attitude maneuvering cannot exceed the energy security range.
	As defined in Equation~\eqref{eq: maneuver}, the initial and ending attitude maneuvering process on each orbit is considered relative to the case where the roll angle is 0.
	Energy constraints~\eqref{Cons4}, namely a relaxation of actual operation \blue{constraints}, contribute to modeling the energy consumption on each orbit and enhancing the energy safety margin.
	\item Constraints~\eqref{Cons5} point out that the value of the decision variable should be restricted as 0 or 1.
	\item Constraints~\eqref{Cons6} \blue{reveal} that the attitude transition time $st_{j}^{l_{1}l_{2}}$ of satellite $j$ should be less than the observation time interval between strips $l_{1}$ and $l_{2}$ on arbitrary orbits $k_{1},k_{2}\in O_{j}$.
	The attitude transition constraint will be verified for the scheduling sequence of each satellite.
\end{itemize}

\section{Greedy Initialization based RPSO Scheduling Algorithm}
\label{sec: IRPSO}

Taking considerable observation strips and the highly combinatorial characteristic of the multi-satellite scheduling problem into account, we have developed an improved RPSO algorithm to solve the large-scale combinatorial optimization problem.
The PSO algorithm is adopted in this study owing to its advantage in searching the strip which contributes most to the overall profit within all available observation strips.
In order to improve the solution efficiently, several modifications in the adaptability of PSO have been proposed.
First, a greedy initialization strategy is integrated to enhance the fast convergence of \blue{the} solution with almost no increase in the calculation.
Then an individual reconstruction method is designed to handle complex operation constraints for generating feasible scheduling solutions.
Furthermore, a resampling mechanism is introduced to the PSO algorithm which will effectively improve the quality of the solution.

\subsection{Framework}

\begin{figure}[htbp]
	\begin{center}
		\includegraphics[width=0.45\textwidth]{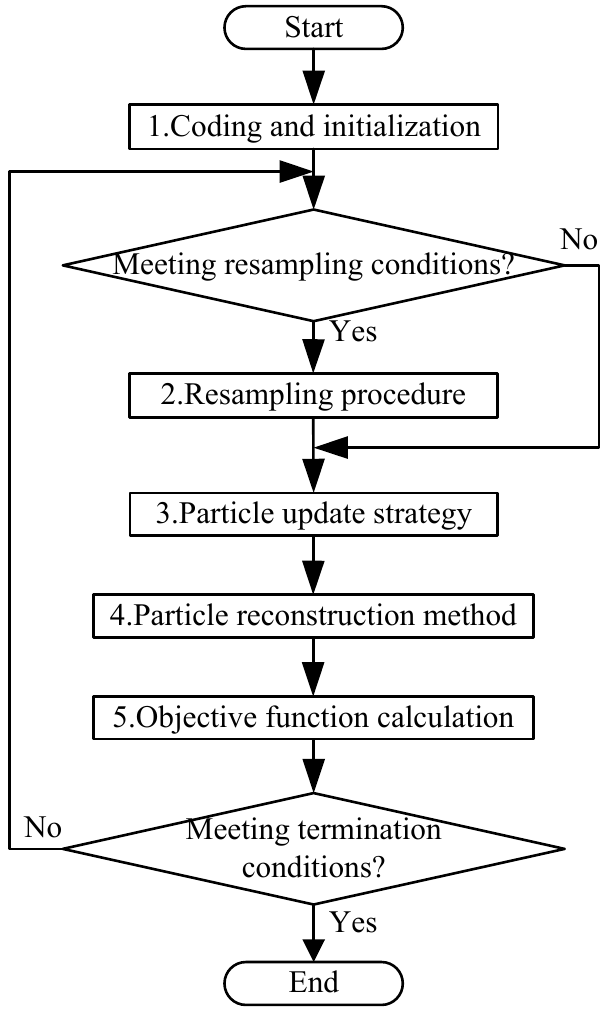}
		\caption{The framework of the scheduling algorithm for large region targets.}\label{fig:GIRPSO}
	\end{center}
\end{figure}

The primary framework of the improved RPSO algorithm is summarized in Figure~\ref{fig:GIRPSO}.
Firstly, each scheduling solution can be described as a particle by introducing the integer coding of particles. 
Meanwhile, most particles are randomly generated except for one particle created based on a greedy strategy.
Subsequently, a resampling procedure will be conducted if the resampling condition is satisfied.
The current position and velocity of particle $n$, namely ${\bm{x}}_n^t$ and ${\bm{v}}_n^t$, will be resampled to overcome the defect of particle hysteresis~\cite{beadle1997fast,liu2001theoretical}. 
According to the PSO algorithm based on the constriction coefficient~\cite{clerc2002particle}, the new velocity $\bm{v}_n^{t + 1}$ and position $\bm{x}_n^{t + 1}$ for each particle $n$ will be updated as follows. 
\begin{equation}
	{\bm{v}}_n^{t + 1} = \chi \left( {{\bm{v}}_n^t + {c_1}{r_1}\left( {{\bm{p}}_n^t - {\bm{x}}_n^t} \right) + {c_2}{r_2}\left( {{\bm{p}}_g^t - {\bm{x}}_n^t} \right)} \right) \label{eq: v}
\end{equation}
\begin{equation}
	\bm{x}_n^{t + 1} = \bm{x}_n^t + \left\lfloor {\bm{v}_n^{t + 1}} \right\rfloor \label{eq: x}
\end{equation}
where $r_{1}$ and $r_{2}$ are random real numbers within [0,1], ${\bm{p}}_n^t$ indicates the current best position of particle $n$ and ${\bm{p}}_g^t$ represents the current global best solution.
Rounding the velocity $\left\lfloor {\bm{v}_n^{t + 1}} \right\rfloor$ could ensure that the particle position remains as an integer.
The constriction coefficient $\chi$ can be determined as
\begin{equation}
	\chi  = \frac{2}{{\left| {2 - \varphi  - \sqrt {{\varphi ^2} - 4\varphi } } \right|}}
\end{equation}
\begin{equation}
	\varphi  = {c_1} + {c_2}, \varphi  > 4
\end{equation}
Generally, $\varphi$ is set as 4.1 while $c_{1}$ and $c_{2}$ are fixed as 2.05.

After the velocity and position \blue{update} process, some particles may not satisfy \blue{the} constraints presented in Section~\ref{subsec: model}. 
In order to ensure that all particles are feasible solutions, a particle reconstruction method is adopted in this study by referring to the individual reconfiguration in~\cite{zhibo2021multi}.
Successively, the objective function~\eqref{ObjFuncNew} will be calculated for each particle based on the coverage calculation method described in Section~\ref{sec: coverage}.
At the same time, the local and global optimum \blue{solutions} will be updated to ${\bm{p}}_n^{t+1}$ and ${\bm{p}}_g^{t+1}$.
Afterward, the termination condition should be verified and the best solution can be output when the stopping criterion is met.
The detailed algorithm framework is described in \blue{the} following sections.

\subsection{Coding and initialization}
\label{subsec: coding}
\begin{figure}[htbp]
	\begin{center}
		\includegraphics[width=0.75\textwidth]{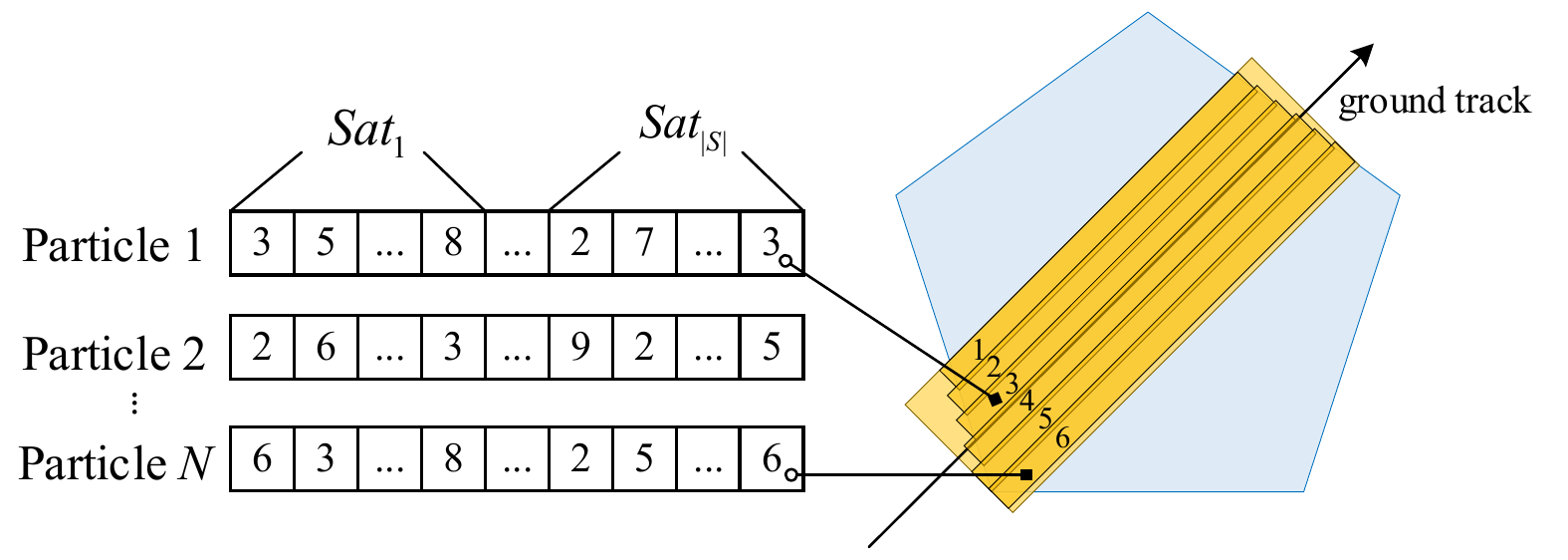}
		\caption{The coding structure of each particle.}
		\label{fig:Coding}
	\end{center}
\end{figure}

Note that the satellite attitude remains unchanged within the observation duration, which means that only one observation strip can be selected during one pass of each EOS.
An integer coding strategy  is adopted to describe each particle, where each gene position corresponds to a visible time window $[VS_{ijk},VE_{ijk}]$.
As shown in Figure~\ref{fig:Coding}, the index value of \blue{each} observation strip is assigned to the corresponding gene.
By arranging according to satellite observations, each particle can represent the scheduling scheme of all satellites.
The number of genes within \blue{a} particle equals the total number of available observation opportunities.
The value of each gene can take an integer varying from 0 to $|L_{ijk}|$, where the zero value indicates that no observation strip is scheduled in this visible time window.

Compared to a completely random initialization method, a greedy initialization strategy for one particle is introduced in our method.
A particle is generated by the greedy strategy while other particles are randomly initialized. 
In \blue{each satellite crossing}, the observation strip with the largest coverage area, corresponding to the maximum observation duration, will be selected \blue{to observe} in the greedy algorithm.
This strategy can guarantee a relatively good initial solution, as well as contribute to the convergence speed of \blue{the} solution.

\subsection{Particle reconstruction method}

\begin{figure*}[htbp]
	\begin{center}
		\includegraphics[width=0.75\textwidth]{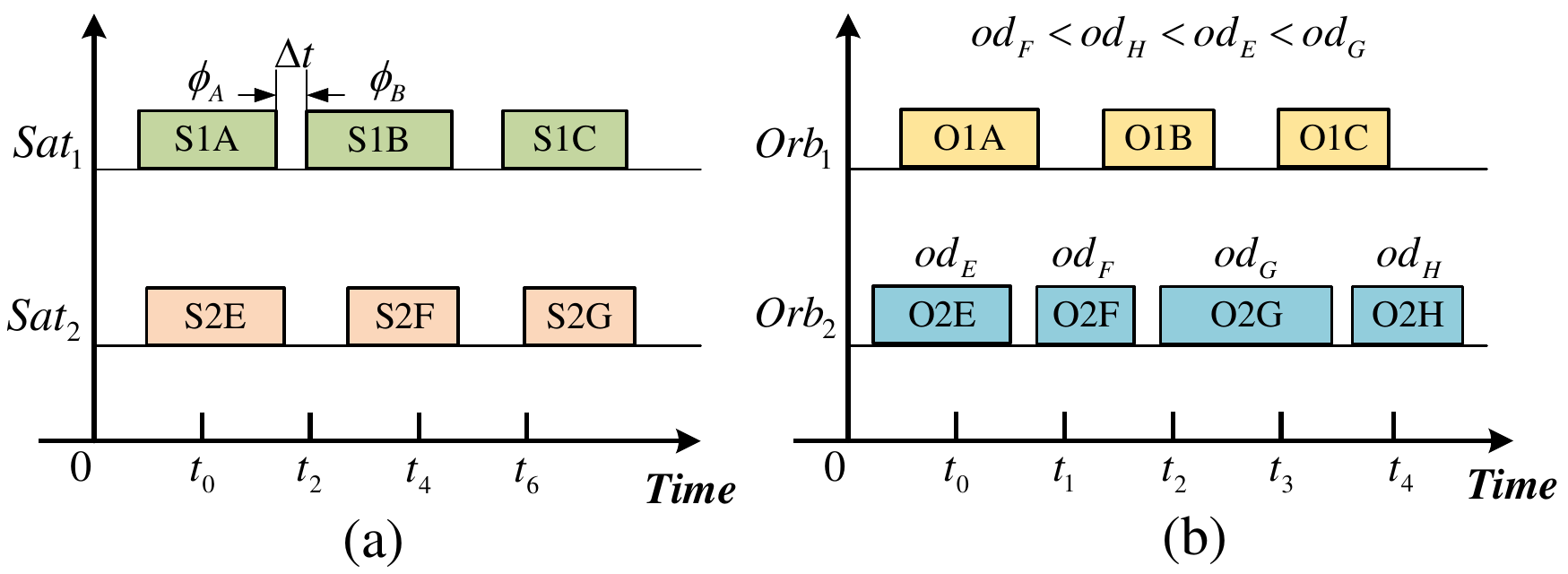}
		\caption{Schematic of particle reconstruction for constraints handling.}
		\label{fig:Plan}
	\end{center}
\end{figure*}

Because constraints~\eqref{Cons1} and~\eqref{Cons5} are already satisfied in the previous heuristic process, there exists no need to handle these constraints in the particle reconstruction procedure. 
For each generated particle, constraints~\eqref{Cons3}-\eqref{Cons4} and~\eqref{Cons6} still should be verified.
As shown in Figure~\ref{fig:Plan}, boxes along the $Time$ axis indicate various observation tasks with different observation durations, where the same color represents that these tasks belong to the same satellite or orbit. 
Attitude transition constraints will be verified for each satellite in subplot (a) while memory and energy constraints should be confirmed on each orbit as depicted in subplot (b).
Note that attitude transition constraints of $Sat_{2}$ are all satisfied owing to \blue{the} sufficient time intervals between adjacent tasks.
With respect to $Sat_{1}$, the time interval $\Delta t$ between observation tasks S1A and S1B is too nervous to accomplish the attitude transition.
In subplot (b), $Orb_{1}$ and $Orb_{2}$ are within the same orbit set $O_{j}$.
Considering the memory and energy consumption, tasks on $Orb_{1}$ can be successfully observed without violating any constraint while $Orb_{2}$ can not.
Therefore, some observation tasks on $Orb_{2}$ should be removed to guarantee all resource constraints.

\begin{algorithm}[htbp]
	\caption{Particle reconstruction algorithm}
	\scriptsize
	\label{al:reconstruction}
	\begin{algorithmic}[1]
		\REQUIRE ~~\\
		Current scheduling solution $Particle_{n}$
		\ENSURE  ~~\\
		The updated scheduling solution
		\FOR{each satellite $j\in S$}
		\WHILE{(1)}
		\STATE Obtain current scheduling scheme of satellite $j\in S$ according to $Particle_{n}$;  
		\STATE Verify the attitude transition constraint for arbitrary two consecutive observation tasks;
		\IF{all constraints are satisfied}
		\STATE break;
		\ELSE
		\FOR{observation strips $\{l_{1},l_{2}\}$ not satisfying constraint~\eqref{Cons6}} 
		\STATE{Adjust the observation strip $l_{2}$ to $l_{2}^{'}$ and update $Particle_{n}$;} \\ 
		\STATE{Verify the attitude transition constraint for the updated $\{l_{1},l_{2}^{'}\}$;}
		\IF{the constraint is not satisfied}
		\STATE Remove the later observation task and update $Particle_{n}$; 
		\ENDIF
		\ENDFOR
		\ENDIF
		\ENDWHILE
		\FOR{each orbit $k \in O_{j}$}
		\STATE{Obtain observation tasks of satellite $j$ on orbit $k$ according to $Particle_{n}$;}
		\STATE{Verify memory constraints~\eqref{Cons3} and energy constraints~\eqref{Cons4} on orbit $k$;} \label{line19}
		\IF{all constraints are satisfied}
		\STATE{break;} \label{line21}
		\ELSE
		\WHILE{(1)}
		\STATE{Remove the task with the minimum observation duration and update $Particle_{n}$;}
		\STATE{Execute Steps~\ref{line19}-\ref{line21};}
		\ENDWHILE
		\ENDIF
		\ENDFOR
		\ENDFOR
	\end{algorithmic}
\end{algorithm}

The \blue{detailed} process of the particle reconstruction method is presented as Algorithm~\ref{al:reconstruction}.
After generating a particle, namely a scheduling solution, the proposed reconstruction algorithm should be conducted for each satellite $j \in S$.
In order to decouple constraints~\eqref{Cons6} and constraints~\eqref{Cons3}-\eqref{Cons4}, the attitude transition constraint will be handled at first and followed by memory and energy constraints.
The transition constraint will be verified for arbitrary two consecutive observation tasks.
Denote $\{l_{1},l_{2}\}$ as two observation strips of two consecutive tasks.
When constraint~\eqref{Cons6} is not satisfied for $\{l_{1},l_{2}\}$, a new strip $l_{2}^{'}$ would be selected.
As can be seen from Figure~\ref{fig:Plan}(a), $\phi_{A}$ remains unchanged while $\phi_{B}$ will be adjusted to make $\Delta t$ as large as possible.
Then the attitude transition constraint will be verified for $\{l_{1},l_{2}^{'}\}$.
The observation task S1B will be removed if constraint~\eqref{Cons6} is still not satisfied.
It is worth noting that whole attitude transition constraints for satellite $j$ will be verified again because the adjustment of \blue{each} task may affect the transition time of adjacent tasks.
The loop will be stopped if all transition constraints are satisfied.

Subsequently, the verification for constraints~\eqref{Cons3} and~\eqref{Cons4} on each orbit $k\in O$ will be conducted.
Both memory and energy consumption are all related to the number of observation tasks and the length of observation duration.
By referring to~\cite{zhibo2021multi}, we will remove the task with the minimum observation duration at first, namely the less coverage.
The observation task on \blue{the} current orbit will be removed in this order until memory and energy constraints are satisfied simultaneously.
Finally, the particle satisfying all constraints can be generated by the reconstruction algorithm.

\subsection{Resampling strategy}
As a core of the proposed improved RPSO algorithm, the resampling strategy contributes to generating a better scheduling solution within given iteration steps.
In the standard PSO algorithm, the particle velocity will be updated as Equation~\eqref{eq: v} and then the new particle position can be generated according to Equation~\eqref{eq: x}.
Generally, the gene value of each newly generated particle position may be out of the valid range $[0, |L_{ijk}|]$. 
When the gene value is larger than $|L_{ijk}|$, it will be fixed as the upper boundary.
Meanwhile, the gene value is set as 0 if a negative number has been assigned to it.
Similarly, each gene value of particle velocity will be limited within the valid range in the same way.

However, there exists a defect in the standard PSO.
Although most particles will converge towards a certain local optimum, some particles may move to several non-optimal regions owing to the random number introduced in the PSO algorithm.
This part of particles contributes little to the optimization process as well as wastes computing resources during the process when they approach better particles.
Therefore, it is necessary to solve the moving hysteresis defect by the resampling strategy.


\begin{figure}[htbp]
	\begin{center}
		\includegraphics[width=0.55\textwidth]{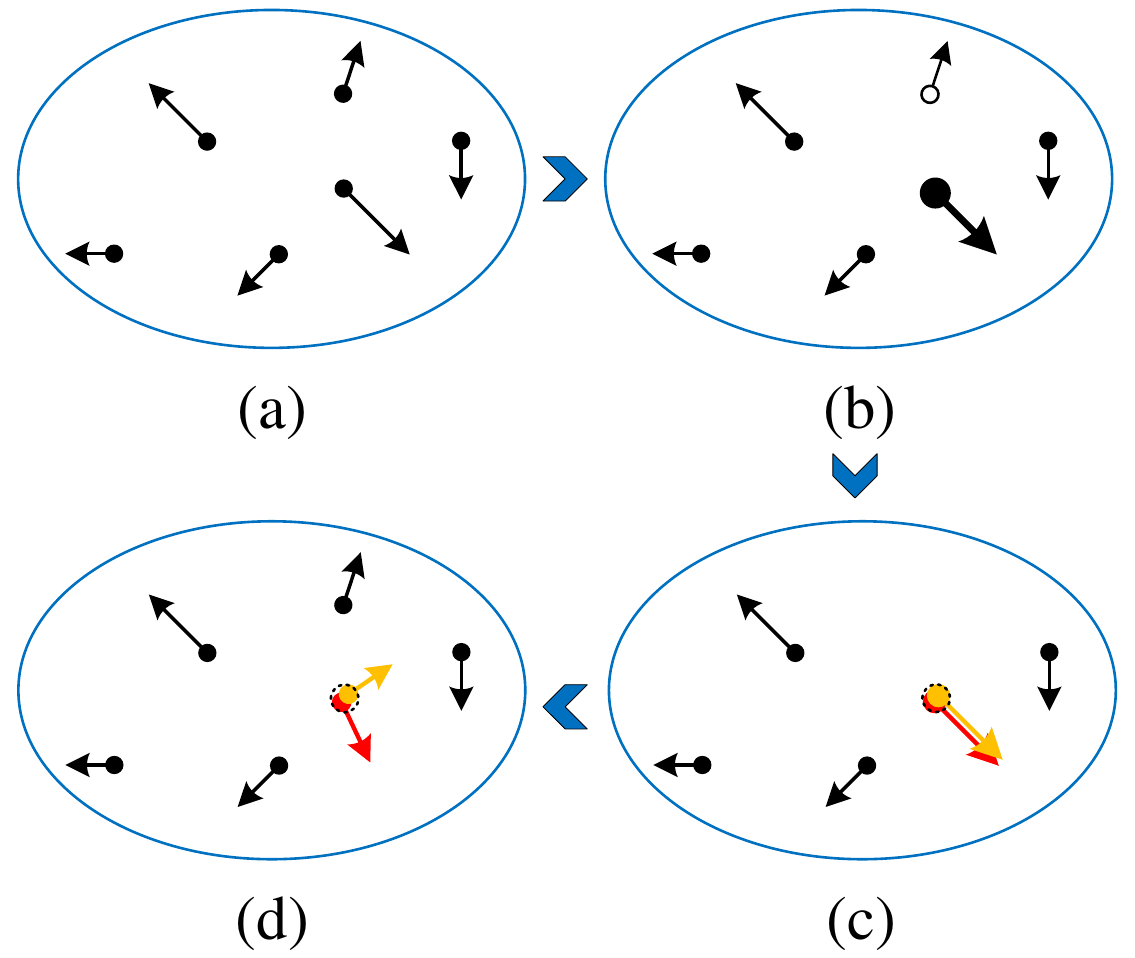}
		\caption{Illustration of the resampling strategy.} 
		\label{fig:Particle}
	\end{center}
\end{figure}


\begin{algorithm}[hbtp]
	\caption{Particle resampling algorithm}
	\scriptsize
	\label{al:resampling}
	\begin{algorithmic}[1]
		\REQUIRE ~~\\
		The position $\bm{x}_{n}^{t}$, velocity $\bm{v}_{n}^{t}$, the objective function value $F(n)$ for each particle $n \in \{1,2,...,N\}$, the current global best solution ${\bm{p}}_g^t$; 
		\ENSURE  ~~\\
		The resampled particle position and particle velocity; 
		\IF{the reampling criteria is met} 
		\STATE Initialization:  $N^{m} \leftarrow 0$; 
		\FOR{each particle $n$}
		\STATE ${q_n} = \dfrac{1}{{\sqrt {2\sigma \pi } }}\exp \left( { - \dfrac{{( {F(n) - {\bm{p}}_g^t} )^2}}{{2\sigma }}} \right)$ 
		\STATE ${Q_n} = q_n/\sum\limits_{n = 1}^N {q_n} $ 
		\ENDFOR
		\STATE $\bar{Q}_n = \dfrac{1}{N} \sum\limits_{n = 1}^{N} {q_n}$; 
		\FOR{$n \in \{1,2,...,N\}$}
		\IF{${Q_n}>\bar{Q}_n$}
		\STATE $N_{c} = \left\lfloor {Q_n/\bar{Q}_n} \right\rfloor$ \\
		\STATE Copy the current particle $N_{c}$ times and set the weight as $\bar{Q}_n$;
		\STATE $N^{m} = N^{m} + N_{c}$ \\
		\ENDIF
		\ENDFOR
		\STATE Inherit $N-N^{m}$ particles from $\{1,2,...,N\}$ and set the weight as $\bar{Q}_n$; 
		\ENDIF
	\end{algorithmic}
\end{algorithm}

As shown in Figure~\ref{fig:Particle}, each dot represents a particle while the arrow indicates the velocity of each particle.
All particles are located in an elliptical area which denotes the feasible region. 
The weight of each particle is calculated and assigned during the process from subplot (a) to (b), where the larger dot indicates the particle with a higher weight and the hollow circle represents the particle with a low weight.
The resampling process is illustrated in the variation from subplot (b) to (c), where the particle with \blue{a} larger weight is copied twice and the particle with a small weight is discarded to keep the total number of particles unchanged.
After resampling, all particle weights are restored to the same level.
Although the newly generated particles are completely the same in Figure~\ref{fig:Particle}(c), the velocity in \blue{the} next time step would be different owing to the random parameters in Equation~\eqref{eq: v}. 
These generated particles will search in different directions from the same location, and then more particles can search for solutions in better areas, which would make up the defect of particle hysteresis to a certain extent.

\blue{The main} steps for \blue{the} resampling algorithm are shown in Algorithm~\ref{al:resampling}.
The resampling criteria here is that the resampling process will be performed twice after every ten iterations by referring to~\cite{wang2018coverage}.
Each particle $n \in \{1,2,...,N\}$ would be given a weight value $q_{n}$ according to its objective function value $F(n)$, which can be determined by Equation~\eqref{ObjFuncNew}.
The parameter $\sigma$ indicates the sample variance of ${F(n) - {\bm{p}}_g^t}$ and $Q_{n}$ represents the unitary weight value of particle $n$.
For each particle $n$, if ${Q_n}>\bar{Q}_n$ is satisfied, particle $n$ will be copied $N_{c}$ times as depicted in Figure~\ref{fig:Particle}(c) and the weight for each particle is set as the average value $\bar{Q}_n$.
The total number of particles obtained by copying is denoted as $N^{m}$.
There are still $N-N^{m}$ particles missing compared to the initial situation.
Finally, these particles will be inherited from original unreplicated particles in descending order of weight.
Particles with smaller weights will be discarded in this procedure.
Through the resampling process, more particles will be concentrated in a better area, which will facilitate the optimization of \blue{the} solution.

\section{Computational Experiments}
\label{sec: result}

Extensive simulations are conducted to evaluate the proposed coverage calculation method and the improved RPSO algorithm in this section.
Three objectives are assigned to the numerical experiments: verify the accuracy of the coverage calculation, evaluate the effectiveness of the improved RPSO algorithm, and conduct the sensitivity analysis. 

\subsection{Data generation}
The scheduling algorithm for large region targets is coded in C++, and experiments are conducted on a desktop with a CPU of 3.20 GHz and a memory of 16.0 GB.
Because there lacks the benchmark for \blue{the} large region targets observation scheduling problem, we have designed the simulation instances by referring to~\cite{xu2020multi,zhibo2021multi}.
The detailed description of \blue{the scenario} is as follows.

\blue{Preliminary simulation shows that the proposed method can achieve the optimal solution in small instances.}
In order to verify the performance of the proposed method \blue{on a practical scale}, seven large region targets distributed in different latitudes of the world are \blue{considered} in the experiment.
The longitude and latitude information of the seven region targets are provided in Table~\ref{tab: Area}, where the area is provided by the System Tool Kit (STK) software according to the geographical location of each region.
These regional targets are assumed to be of equal priority in the observation requirement from customers.

\begin{table*}[htbp]
	\centering
	\footnotesize
	\caption{Basic information of seven large region targets.}
	\begin{tabular}{cccccr}
		\toprule
		Targets & \multicolumn{4}{c}{Latitude \& Longitude} & \multicolumn{1}{c}{Area($km^{2}$)} \\
		\midrule
		$T_{1}$ & \multicolumn{4}{c}{(5.33,30.26),(5.33,9.32),(-5.17,9.32),(-5.17,30.26)} & 2732840.4  \\
		$T_{2}$ & \multicolumn{4}{c}{(-0.53,-50.5),(-0.53,-75.26),(-9.06,-75.26),(-9.06,-50.5)} & 2628413.6  \\
		$T_{3}$ & \multicolumn{4}{c}{(32.58,114.91),(32.58,105.09),(22.09,105.09),(22.09,114.91)} & 1129008.3  \\
		$T_{4}$ & \multicolumn{4}{c}{(20.557,111.658),(10.52,111.658),(10.52,119.35),(20.557,119.355)} & 916658.2  \\
		$T_{5}$ & \multicolumn{4}{c}{(34.67,178.48),(24.44,178.48),(24.44,168.15),(34.67,168.15)} & 1134530.3  \\
		$T_{6}$ & \multicolumn{4}{c}{(35.5,85.5),(25.8,85.5),(25.8,95.5),(35.5,95.5)} & 1030102.7  \\
		$T_{7}$ & \multicolumn{4}{c}{(70.5,21.5),(60.8,21.5),(60.8,33.8),(70.5,33.8)} & 608088.1  \\
		\bottomrule
	\end{tabular}%
	\label{tab: Area}%
\end{table*}%

The simulation experiment is conducted from the same start time of 2021-04-07 00:00:00. 
Twenty satellites in the Chinese satellite platform are selected, and initial orbital parameters and the maximal roll angle $\phi_{max}$ of involved EOSs are shown in Table~\ref{tab: Sat}.
Meanwhile, other performance parameters of satellites and orbits are set as the same and have been shown in Table~\ref{tab: parameters}.

\begin{table*}[htbp]
	\centering
	\footnotesize
	\caption{Specific orbital elements and maneuvering ability of the selected satellites.}
	\begin{tabular}{rrrrrrrc}
		\toprule
		\multicolumn{1}{c}{$ID$} & \multicolumn{1}{c}{$a$(km)} & \multicolumn{1}{c}{$e$} & \multicolumn{1}{c}{$i$($^{\circ}$)} & \multicolumn{1}{c}{$\Omega$ ($^{\circ}$)} & \multicolumn{1}{c}{$\omega$ ($^{\circ}$)} & \multicolumn{1}{c}{$M$ ($^{\circ}$)} & $\phi_{max}$($^{\circ}$) \\
		\midrule
		Sat1  & 7015.341  & 0.001108  & 97.8860  & 339.1370  & 155.8560  & 20.9230  & 30  \\
		Sat2  & 7002.399  & 0.001291  & 97.8560  & 54.1270  & 157.1140  & 306.0990  & 35  \\
		Sat3  & 7126.453  & 0.001121  & 98.5290  & 91.1400  & 90.2280  & 268.9880  & 30  \\
		Sat4  & 7015.610  & 0.000861  & 97.9960  & 140.7450  & 163.0120  & 219.4300  & 30  \\
		Sat5  & 7149.376  & 0.001479  & 98.7240  & 117.3840  & 140.6560  & 242.8550  & 32  \\
		Sat6  & 6876.495  & 0.001410  & 97.3820  & 96.3900  & 155.0710  & 56.6480  & 30  \\
		Sat7  & 6997.446  & 0.002413  & 97.8490  & 215.1140  & 109.3060  & 186.4790  & 30  \\
		Sat8  & 7001.883  & 0.001245  & 97.9200  & 235.5390  & 90.2690  & 114.1530  & 30  \\
		Sat9  & 7001.856  & 0.001260  & 97.8600  & 229.4870  & 90.0140  & 354.3360  & 30  \\
		Sat10 & 6869.771  & 0.001486  & 97.3560  & 350.2770  & 77.9030  & 271.3230  & 30  \\
		Sat11 & 6887.457  & 0.001387  & 97.5150  & 161.0020  & 90.5810  & 271.2620  & 30  \\
		Sat12 & 6886.176  & 0.001263  & 97.3050  & 300.6480  & 87.0550  & 9.4970  & 30  \\
		Sat13 & 6887.253  & 0.001410  & 97.4430  & 170.5300  & 87.5400  & 79.7450  & 30  \\
		Sat14 & 7618.359  & 0.002772  & 99.9940  & 240.6510  & 47.7570  & 255.9060  & 30  \\
		Sat15 & 7578.863  & 0.000758  & 100.4130  & 7.0110  & 60.8300  & 337.8330  & 30  \\
		Sat16 & 7579.543  & 0.000695  & 100.2430  & 301.4140  & 79.4160  & 138.3540  & 30  \\
		Sat17 & 7578.169  & 0.000451  & 100.3980  & 345.9620  & 75.0290  & 250.0000  & 30  \\
		Sat18 & 7575.721  & 0.001826  & 100.3020  & 293.3490  & 77.6880  & 332.1050  & 30  \\
		Sat19 & 7142.139  & 0.000764  & 98.3110  & 301.8740  & 186.2800  & 54.1290  & 30  \\
		Sat20 & 7142.139  & 0.000780  & 101.3580  & 223.8170  & 82.2040  & 23.1210  & 30  \\
		\bottomrule
	\end{tabular}%
	\label{tab: Sat}%
\end{table*}%

\begin{table}[htbp]
	\centering
	\footnotesize
	\caption{Constraint parameters for each satellite and orbit.}
	\begin{tabular}{cc}
		\toprule
		Parameter & Value \\
		\midrule
		$M_{jk}$ & 30000 MB \\
		$mo_{jk}$ & 150 MB/s \\
		$E_{jk}$ & 40000 J \\
		$eo_{jk}$ & 100 J/s \\
		$et_{jk}$ & 150 J/$^{\circ}$ \\
		$v_{j}^{roll}$ & 3 $^{\circ}$/s \\
		\bottomrule
	\end{tabular}%
	\label{tab: parameters}%
\end{table}%

The particle population size generally varies from 20 to 50 in practical applications~\cite{clerc2002particle,wang2018coverage}, and we set the particle size as 50.
According to Section~\ref{subsec: coding}, the position range of each gene value is [0, $L_{ijk}$], and the corresponding velocity range is set as [-$\left\lfloor |L_{ijk}|/10 \right\rfloor$, $\left\lfloor |L_{ijk}|/10 \right\rfloor$].
The parameter $\delta \phi=0.001 rad$ is initialized to obtain discrete observation strips.

\subsection{Results and discussions}

\subsubsection{Coverage calculation}
\begin{figure}[htbp]
	\begin{center}
		\includegraphics[width=0.75\textwidth]{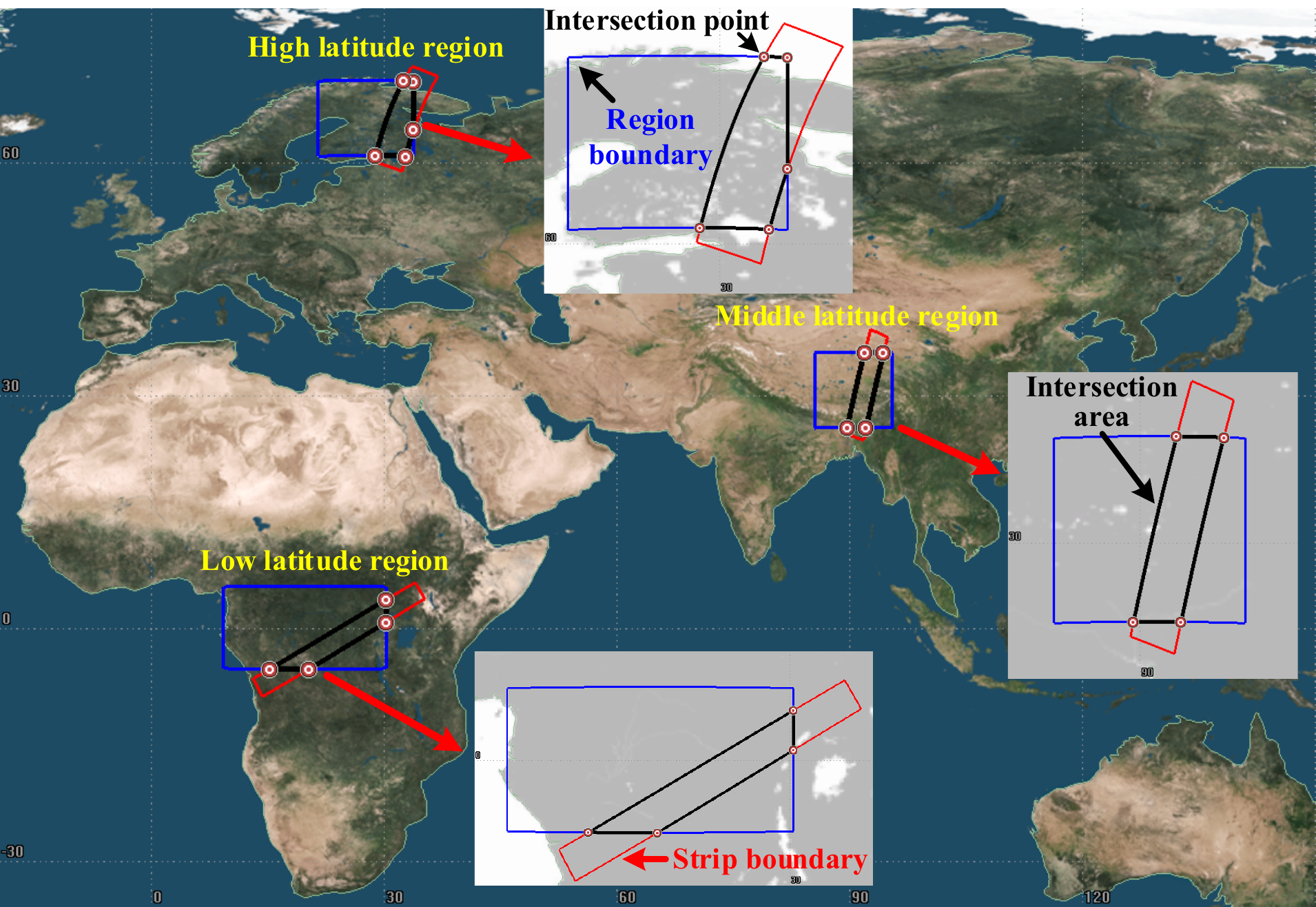}
		\caption{Three large regions at different latitudes for coverage calculation comparison.} 
		\label{fig:ThreeArea}
	\end{center}
\end{figure}

First, the proposed grid-less coverage calculation method in Section~\ref{sec: coverage} is verified by comparing the GDAC method and the STK software.
As shown in Figure~\ref{fig:ThreeArea}, large region targets and observation strips located at different latitudes are selected in STK for the comparison experiment.
Areas enclosed by the blue border are three large region targets from low latitude to high latitude, which correspond to $T_{1}$, $T_{6}$, and $T_{7}$ in Table~\ref{tab: Area}, respectively.
Observation strips are represented as the area enclosed by the red border and can be generated arbitrarily to verify the coverage calculation result.
All intersection points between observation strips and region targets can be obtained accurately by the \blue{APCT} in this paper.
As depicted in Figure~\ref{fig:ThreeArea}, we have added the intersection point to STK according to the latitude and longitude information computed by APCT.
It is not difficult to find that these intersection points are exactly located on the region boundary, which indicates that intersection calculation results are consistent with \blue{the} two-dimensional display results of STK.
The intersection area consists of black boundaries connecting intersection points, and then the coverage calculation can be compared.

The coverage can be directly obtained based on the APCT in this paper.
According to the geographical location information of the large region target and the intersection area, STK can provide the corresponding acreage value, and then the coverage can be calculated.
The GDAC method would count the number of discrete grid points within the intersection area. 
The different dispersion of grid points can be adjusted by STK and \blue{the} coordinates of each point are output for the counting by GDAC. 
Comparison results are shown in Table~\ref{tab: coverage}, where column Dis indicates the different dispersion of \blue{the} grid and column Cov indicates \blue{the} coverage calculated by different methods.
What stands out in the table is that the absolute error between APCT and STK is exclusively small in all three instances, which are completely negligible in practical engineering.
The program running time for almost all cases by GDAC is higher than that of APCT in addition to one result highlighted by the underscore.
Meanwhile, the coverage calculation result by GDAC is throughout not as accurate as APCT, although its running time has exceeded 1000 times longer than APCT when dispersion equals 0.05$^{\circ}$.
This might be explained by the inherent defect of GDAC that grid points close to the boundary are not completely inside or outside the area. %
Then the coverage calculation result is accurate only when the dispersion is small enough, which results in an exhaustively increase in running time.
Closer inspection of Table~\ref{tab: coverage} shows the relative errors between GDAC and STK are 1.91\%, 6.62\%, and 9.90\% for the three instances when the dispersion is 1.0$^{\circ}$.
This indicates that the coverage calculation error is expected to decline as the region area increases under the same degree of dispersion.
In conclusion, the presented APCT in this paper can calculate the coverage accurately within a shorter time, which is definitely an improvement to the calculation method by grid points.

\begin{table*}[htbp]
	\centering
	\footnotesize
	\caption{Comparison of coverage calculation results of different methods.}
	\begin{tabular}{cccrrcrrcr}
		\toprule
		\multirow{2}[4]{*}{Method} & \multicolumn{1}{r}{\multirow{2}[4]{*}{Dis/$^{\circ}$}} & \multicolumn{2}{c}{Low latitude region} &       & \multicolumn{2}{c}{Middle latitude region} &       & \multicolumn{2}{c}{High latitude region} \\
		\cmidrule{3-4}\cmidrule{6-7}\cmidrule{9-10}          &       & Cov/$\%$ & \multicolumn{1}{c}{Time/$ms$} &       & Cov/$\%$ & \multicolumn{1}{c}{Time/$ms$} &       & Cov/$\%$ & \multicolumn{1}{c}{Time/$ms$} \\
		\midrule
		GDAC  & 1.000  & 16.4021  & 4173.00  &       & \uline{26.3889}  & \uline{493.00}  &       & 28.8889  & 1210.00  \\
		& 0.500  & 16.4903  & 25189.00  &       & 23.8683  & 3390.00  &       & 27.0370  & 3780.00  \\
		& 0.100  & 16.7409  & 344398.00  &       & 24.7928  & 61455.00  &       & 26.2234  & 35950.00  \\
		& 0.050  & 16.7490  & 1962054.00  &       & 24.6708  & 528109.00  &       & 26.2438  & 127513.00  \\
		STK   & $-$   & 16.7219  & $-$   &       & 24.7493  & $-$   &       & 26.2875  & $-$ \\
		APCT  & $-$   & 16.7217  & 1186.00  &       & 24.7494  & 600.00  &       & 26.2908  & 631.00  \\
		\bottomrule
	\end{tabular}%
	\label{tab: coverage}%
\end{table*}%

\subsubsection{Large region targets observation scheduling}
\begin{figure}[htbp]
	\begin{center}
		\includegraphics[width=0.62\textwidth]{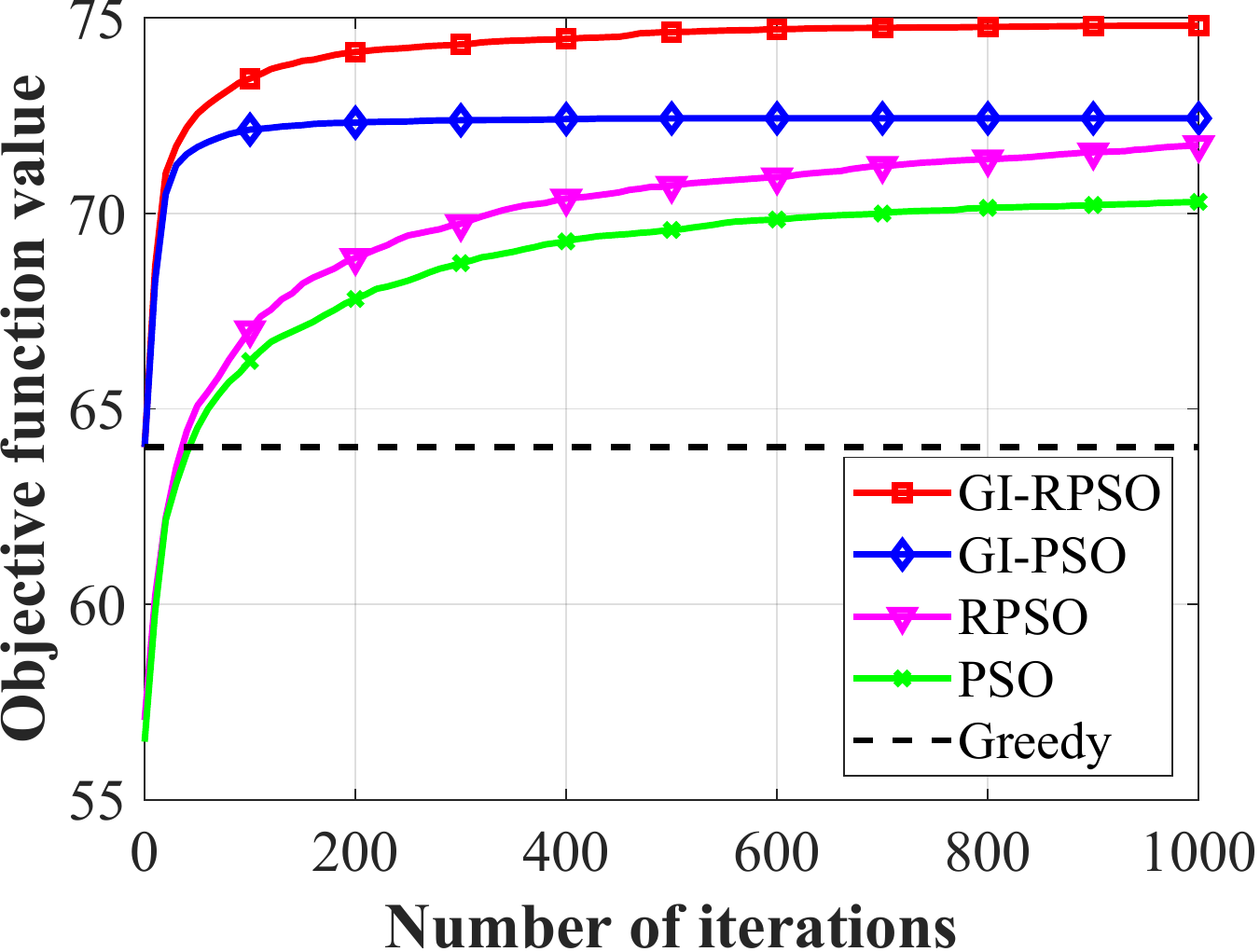}
		\caption{Performance comparison of several algorithms at different iterations.} 
		\label{fig:Resample}
	\end{center}
\end{figure}

The performance of the improved greedy initialization-based RPSO (GI-RPSO) algorithm is evaluated by comparing the greedy initialization-based PSO (GI-PSO) algorithm, \blue{the} RPSO algorithm with no greedy strategy, and the standard PSO algorithm.
The scenario duration is set as 1 day and the maximal iteration number is initialized as 1000 to validate the algorithm performance.
Without loss of generality, 10 runs for each algorithm have been conducted and the average objective function value of Equation~\eqref{ObjFuncNew} within the iteration can be calculated and depicted in Figure~\ref{fig:Resample}.
Owing to the introduced greedy initialization strategy, the initial solution of GI-RPSO and GI-PSO is the same as the greedy result.
It is apparent that there is a sharp rise in the scheduling result for all four algorithms with the increase \blue{in} iteration numbers.
Though the random initial solutions of PSO and RPSO are much worse than the greedy result, an opposite situation would arise when the iteration number exceeds 100. 
What is interesting in this figure is the rapid convergence of the scheduling result of GI-RPSO and GI-PSO, which could be attributed to the introduction of the greedy strategy.
From the data point in Figure~\ref{fig:Resample}, it is apparent that the resampling method could contribute to the improvement of the scheduling solution.
Average objective function values of GI-RPSO and RPSO are consistently higher than that of GI-PSO and PSO, respectively.
Closer inspection of the figure shows that the result of GI-PSO would converge quickly, which would be trapped in the local optima.
A higher objective function value can be obtained by GI-RPSO because the resampling strategy could contribute to searching for a better solution.
Considering that the greedy algorithm is widely utilized in practical engineering~\cite{xu2020multi}, the comparison experiment with the greedy method can validate the efficiency of the proposed algorithm in this study.
Meanwhile, the GI-RPSO performs the best among all five algorithms, which emphasizes the ability of the improved RPSO algorithm in the large region targets observation scheduling problem.
It can be concluded that the proposed resampling method and greedy initialization strategy perform well in the improved RPSO algorithm.

\begin{figure}[htbp]
	\begin{center}
		\includegraphics[width=0.65\textwidth]{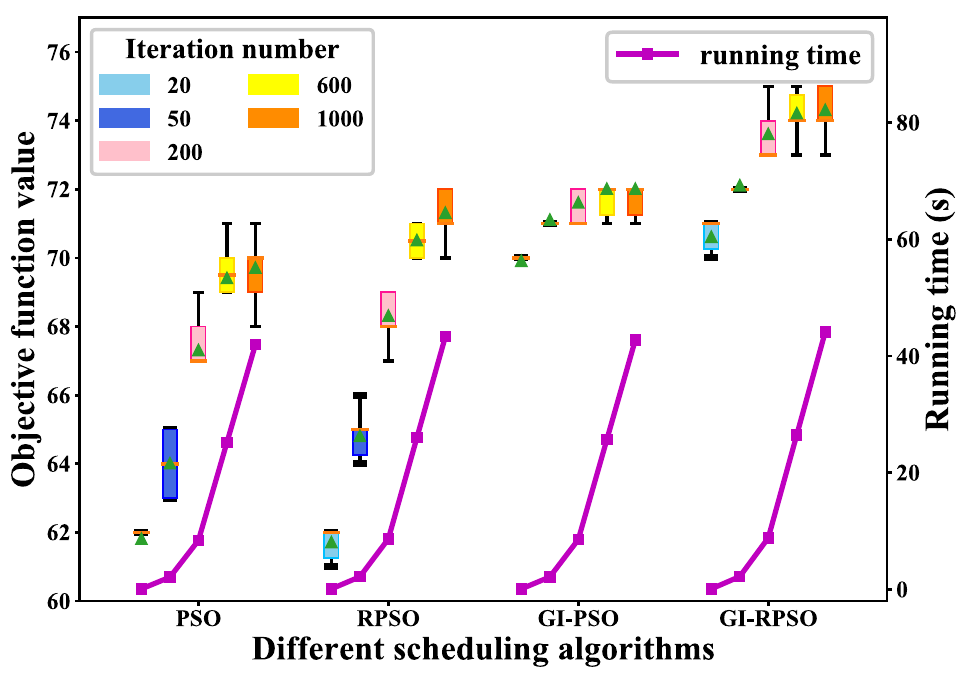}
		\caption{The comparison result of different algorithms in different iterations.} 
		\label{fig:objective}
	\end{center}
\end{figure}

The distribution of scheduling results of 10 runs at different iteration numbers is shown in Figure~\ref{fig:objective}, where the green triangle symbol and the orange horizontal line in the box represent the average and the median of 10 results, respectively.
Each box indicates the distributions, and different surface colors represent various iteration numbers in line with the legend.
The program execution time of the scheduling algorithm for each run is denoted as the purple square dot.
Overall, there exists no big difference in scheduling results among different runs.
Besides, the introduction of the greedy strategy might enhance the stability of the algorithm, which could be found by the comparison of scheduling results between PSO and GI-PSO.
It is not difficult to find that \blue{the} running times of different algorithms at the same iteration number are analogous, which indicates that proposed strategies are not time-consuming.
The program running time increases with the growth of iteration numbers, and the maximal running time does not exceed 50 seconds, which shows the excellent efficiency of the proposed algorithm.

\begin{figure}[htbp]
	\begin{center}
		\includegraphics[width=0.65\textwidth]{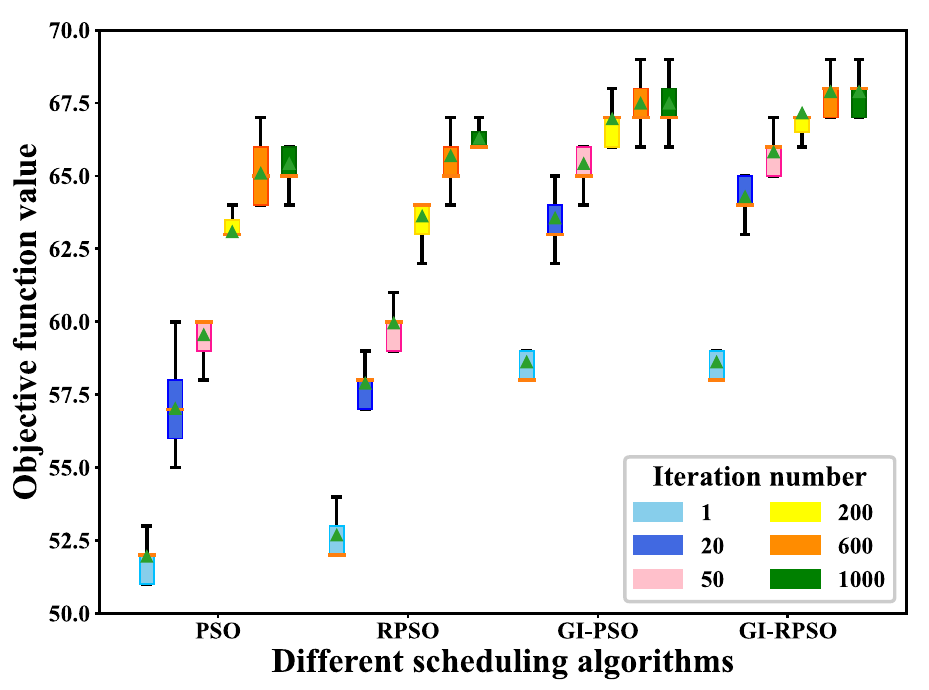}
		\caption{\blue{The statistical result of various target sets in different algorithms.}}
		\label{fig:targetset}
	\end{center}
\end{figure}

\blue{To further evaluate the performance of various algorithms, fifteen unique target sets are generated randomly to conduct the comparison experiment.
Each target set contains fifteen regions, including the previous seven regions and eight newly generated targets.
In line with~\cite{he2020balancing}, these eight targets are randomly generated as a square region with a width of 4$^{\circ}$, located within the latitude range [60$^{\circ}$S, 60$^{\circ}$N] and the longitude range [180$^{\circ}$W, 180$^{\circ}$E].
For each target set, 10 runs are conducted and the average results in different iterations are recorded.
The statistical results of fifteen target sets are presented in Figure~\ref{fig:targetset}.
We can observe a steady increase in the objective function value as the iteration number grows with respect to all four algorithms.
The performance of the greedy algorithm is provided by the results of the first iteration of GI-PSO and GI-RPSO.
Note that the PSO and RPSO will exceed the greedy algorithm after about 50 iterations.
The results of GI-PSO and GI-RPSO at 200 iterations are better than the ultimate performance of PSO and RPSO, demonstrating the promotion effect of the greedy-initialization strategy.
Meanwhile, it can be seen that the average results of GI-RPSO outperform other algorithms, which validates the effectiveness of the proposed heuristic.
Considering the performance of GI-RPSO improves little when the iteration number varies from 600 to 1000, the maximal iteration will be set as 600 in the following experiments.}

The verification for the proposed particle reconstruction method is conducted by experiments with different constraint parameters.
As shown in Table~\ref{tab:memory}, the value of $mo_{jk}$ can be adjusted and corresponding simulation results by the GI-RPSO algorithm in different iterations are recorded.
\blue{Parameters} $mo_{jk}$ \blue{vary} from 100 to 200, and column \blue{IFN} indicates the initial feasible number of particles.
It can be found that for the instance with $mo_{jk}=100$MB/s, all \blue{initially} generated particles are feasible and the reconstruction is not performed.
When the value of $mo_{jk}$ increases to 125, there exist 7 initial particles that do not meet constraints.
Meanwhile, the particle generated by the greedy strategy is still feasible in this instance according to the simulation result at the first iteration.
If $mo_{jk}$ exceeds 150 MB/s, however, there exists no feasible particle at the initialization procedure.
Then all particles will be changed to the solution satisfying all constraints by utilizing the particle reconstruction method.
What can be clearly seen in Table~\ref{tab:memory} is the steady decline of the objective function value at each iteration with the increase of $mo_{jk}$.
There is a dramatic decline in the average value of the objective function when $mo_{jk}$ varies from 150 to 200, which might be explained by the fact that the energy constraints~\eqref{Cons3} would be violated seriously in these instances.
It can be concluded that the constraint is evaluated in this experiment and the proposed particle reconstruction method performs efficiently.

\begin{table*}[htbp]
	\centering
	\footnotesize
	\caption{Comparison of simulation results under different constraint parameters.}
	\begin{tabular}{ccccccccc}
		\toprule
		\multirow{2}[4]{*}{\boldmath{}\textbf{$mo_{jk}$}\unboldmath{}} & \multirow{2}[4]{*}{\blue{\textbf{IFN}}} & \multicolumn{7}{p{29.33em}}{\textbf{Simulation results in different iteration numbers}} \\
		\cmidrule{3-9}          &       & 1     & 100   & 200   & 300   & 400   & 500   & 600 \\
		\midrule
		100   & 50    & 71.574  & 82.704  & 83.555  & 83.719  & 83.880  & 83.973  & 84.007  \\
		125   & 43    & 71.574  & 82.145  & 82.935  & 83.186  & 83.325  & 83.368  & 83.392  \\
		150   & 0     & 64.022  & 73.449  & 74.128  & 74.326  & 74.466  & 74.635  & 74.712  \\
		175   & 0     & 39.343  & 47.620  & 49.119  & 50.076  & 50.620  & 51.072  & 51.390  \\
		200   & 0     & 18.485  & 25.608  & 26.757  & 27.458  & 27.887  & 28.223  & 28.446  \\
		\bottomrule
	\end{tabular}%
	\label{tab:memory}%
\end{table*}%

\begin{figure}[htbp]
	\begin{center}
		\includegraphics[width=0.65\textwidth]{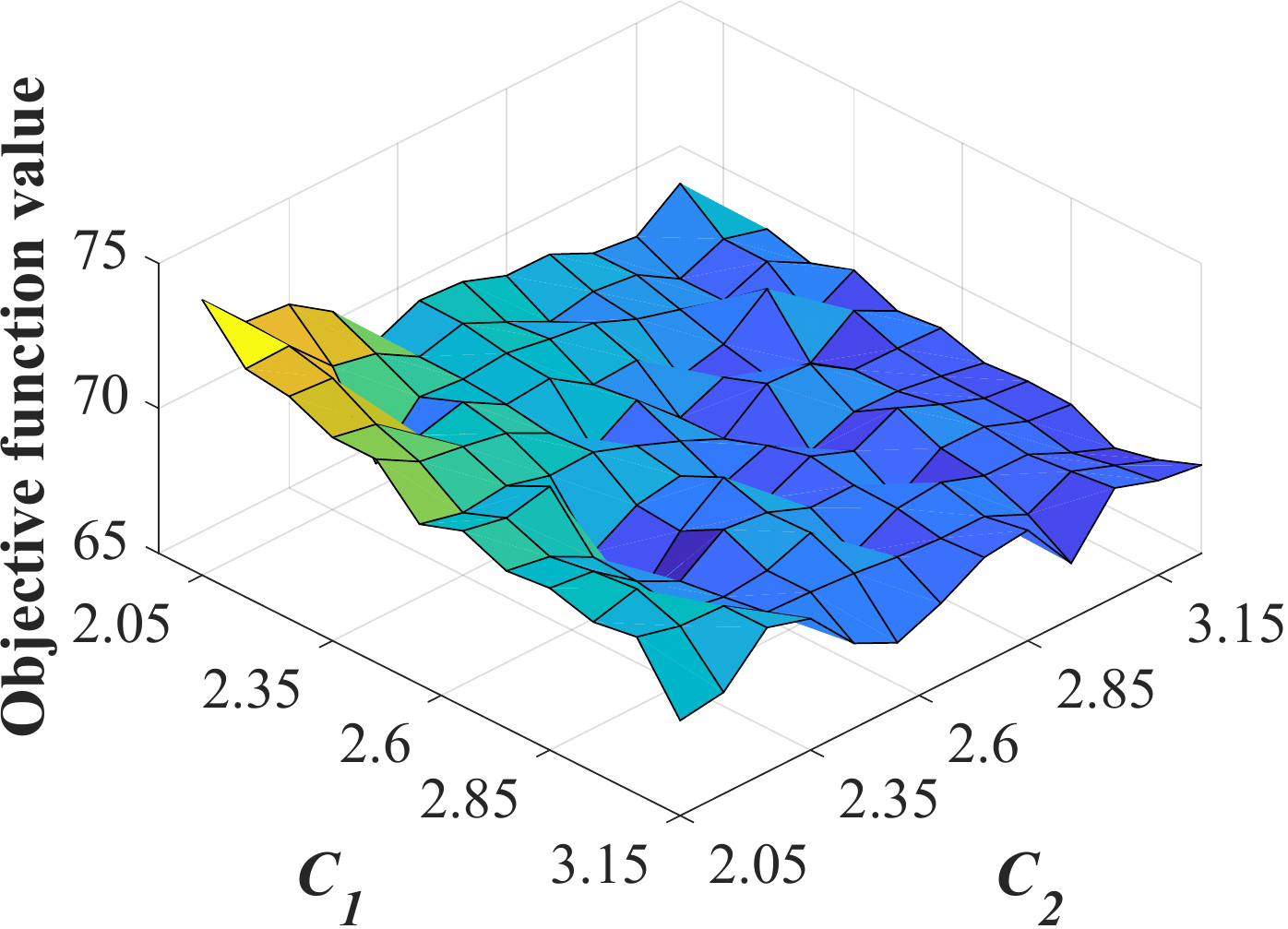}
		\caption{\blue{Sensitivity analysis of coefficients $c_1$ and $c_2$.}}
		\label{fig:Surface}
	\end{center}
\end{figure}

\blue{The coefficients $c_{1}$ and $c_{2}$ in Equation~\eqref{eq: v} could impact the optimization result of the proposed heuristic.
To quantify the impact of $c_{1}$ and $c_{2}$, a sensitivity analysis has been conducted with respect to the proposed GI-RPSO algorithm. 
The seven large regions are scheduled to observe by the selected 20 satellites, and the number of iterations is set as 600.
Following~\cite{wang2018coverage}, $c_{1}$ and $c_{2}$ are initially both set as 2.05 to satisfy the constraint that the sum of $c_{1}$ and $c_{2}$ exceeds 4.
We then vary two coefficients from 2.05 to 3.15, with a step of 0.1 for the experiment.
For each combination of $c_{1}$ and $c_{2}$, ten runs are conducted.
Therefore we entirely have $12 \times 12 \times 10 = 1440$ runs.
The simulation results are shown in Figure~\ref{fig:Surface}, where the objective function value is reported by a surface plot with the horizontal axes $c_{1}$ and $c_{2}$.
Clearly the objective function value has a decreasing trend with the growth of $c_{1}$ and $c_{2}$, although results fluctuate in a few cases.
It can be observed that the highest objective function value is obtained when $c_{1}$ and $c_{2}$ are both set as 2.05, which is consistent with the recommended values in~\cite{clerc2002particle}.}

\begin{figure}[htbp]
	\begin{center}
		\includegraphics[width=0.65\textwidth]{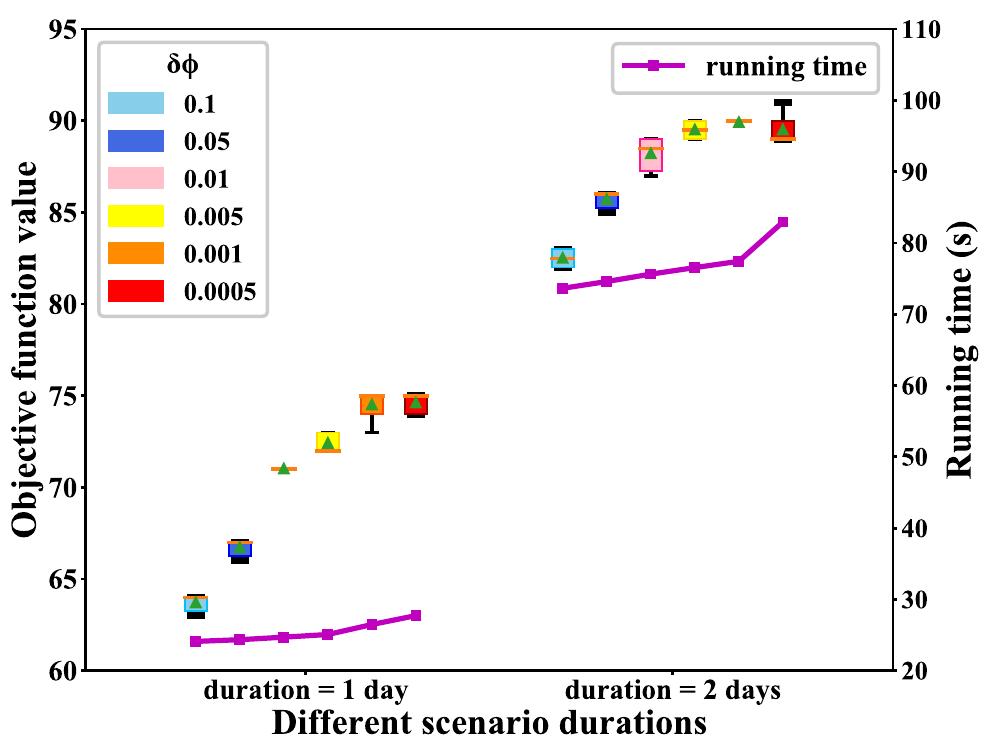}
		\caption{Sensitivity analysis of $\delta \phi$ under different scenario durations.} 
		\label{fig:dPhi}
	\end{center}
\end{figure}

In order to evaluate the influence of discrete granularity of \blue{the} roll angle, \blue{a} sensitivity analysis of $\delta \phi$ has been conducted.
Simulation results under two different scenario durations are shown in Figure~\ref{fig:dPhi}, where the iteration number is taken as 600.
Different values of $\delta \phi$ varying from 0.1 to $0.0005 rad$, as depicted in the legend, are chosen for the comparative experiment.
What is interesting in this figure is the steady growth of the objective function value with the decrease of $\delta \phi$. 
This result may be explained by the fact that more available observation strips can be provided when the value of $\delta \phi$ is set as a smaller value. 
The increased solution space may potentially contain better scheduling results, and then the improved RPSO could search for a better solution compared to the instance with a higher value of $\delta \phi$.
Figure~\ref{fig:dPhi} shows that the difference of results under $\delta \phi$ equals 0.001 and 0.0005 is not obvious, which indicates that $\delta \phi=0.001 rad$ is fine enough to obtain an excellent scheduling result.
Therefore the value of $\delta \phi$ is set as $0.001 rad$, namely about 0.057$^{\circ}$, for \blue{the} corresponding experiments in this study.
Meanwhile, it is not difficult to find that the program running time changes insignificantly with the variation of $\delta \phi$, which indicates the excellent efficiency of the proposed scheduling algorithm. 
According to the data provided \blue{on} the website of Natural Resources Satellite Remote Sensing Cloud Service Platform~\cite{sasclouds.org}, the attitude pointing accuracy of the high-resolution multi-mode satellite is better than 0.01$^{\circ}$.
Owing to the fact that a considerable degree of pointing accuracy can be achieved by \blue{a} practical satellite, the parameter value set in this article \blue{is} reasonable.

\section{Conclusions}
\label{Sec: conclu}
We have addressed the large region targets observation scheduling problem by multiple EOSs in this work. 
An efficient coverage calculation method has been developed based on the polygon clipping technique.
Meanwhile, an improved RPSO scheduling algorithm has been proposed to solve the multiple large region targets scheduling problem, where the greedy initialization strategy and the resampling method perform significant roles.
Comparative experiments show that the proposed coverage calculation method with respect to the large region exhibits better performance than the GDAC method in terms of calculation accuracy and efficiency.
This approach will be useful in improving the calculation ability of large region coverage problems in practical engineering.
In the scheduling process, the greedy initialization strategy has been proven effective in enhancing the convergence speed and solution quality of the optimization algorithm.
The resampling method contributes to searching for a better solution, and the proposed GI-RPSO outperforms other heuristics, especially the greedy algorithm widely used in practice.
Extensive experiments have been conducted to verify the effectiveness and stability of the proposed algorithm.   

Further studies regarding the role of agile EOSs in the large region targets observation would be worthwhile.
Compared to the non-agile EOSs,  agile satellites possess stronger attitude maneuverability, leading to more observation opportunities. In addition, the cloud occlusion to optical sensors could be taken into consideration for future research.

\normalem
\bibliographystyle{IEEEtran}
\bibliography{SatScheduling}

\end{document}